\documentclass[preprint,12pt,authoryear]{elsarticle}

\usepackage{graphicx}
\usepackage{amsmath}
\usepackage{xspace}
\usepackage{lineno}
\usepackage{ifthen}
\usepackage{lscape}
\usepackage{placeins}
\usepackage{color}
\usepackage{epstopdf}
\usepackage{amssymb}

\journal{Icarus}

\begin{document}

\begin{frontmatter}



\title{The 2~$\mu$m spectrum of the auroral emission in the polar regions of Jupiter}


\author[lable1]{L. Kedziora-Chudczer}
\ead{lkedzior@unsw.edu.au}
\author[lable1]{D. V. Cotton}
\author[lable2]{D. J. Kedziora}
\author[lable1]{J. Bailey}

\address[lable1]{School of Physics, UNSW, 2052, Sydney, NSW Australia}
\address[lable2]{School of Physics, The University of Sydney, 2006, Sydney, NSW Australia}

\begin{abstract} 
We report observations of the high (R~$\sim$~18000) and medium (R~$\sim$~5900)
resolution, near-infrared spectra of Jupiter's polar regions with the GNIRS instrument at the Gemini North telescope. \color{black}The observations correspond to the area of main auroral oval in the South and the main spot of the Io footprint in the North. \color{black} We
detected and assigned 18 emission lines of the H$_{3}^{+}$, 2$\nu_{2}\rightarrow 0$ overtone band in the region from 4800
to 4980 cm$^{-1}$ and 5 additional lines in the extended low-resolution spectrum. We use our new modelling scheme, ATMOF to
remove telluric absorption bands of CO$_2$ that feature strongly in the 2~$\mu$m region. The H${_2}$ 1-0 S(1), S(2) and S(3)
emission lines are also detected in the observed spectral region. We found the rotational temperature and column density of
H$_{3}^{+}$ emission at the peak intensity for both northern and southern auroral regions to be the same within the
measurement errors (T$_{rot} \sim~950$K and N(H$_{3}^{+}$) $\sim$ 4.5$\times10^{16}$~m$^{-2}$). The estimates of T$_{rot}$
from H${_2}$ are consistent within much higher uncertainties with temperatures derived from H$_{3}^{+}$ emissions. 
\color{black}We derived the profiles of the H$_{3}^{+}$ emissivity and ion density for both auroral regions providing the first such measurement for the emission associated with the main spot of the Io footprint. 
\color{black} We also found
a number of weaker lines in the high-resolution spectra that could be associated with emission from high excitation levels
in neutral iron, which could be deposited in Jupiter's atmosphere as a result of meteor ablation.

\end{abstract}


\end{frontmatter}







\section{Introduction} \label{sec1} Jupiter's auroral emissions were first observed in the ultraviolet by Voyager 1 in 1979.
Since then the Jovian aurora has been detected from space in X-rays by the ROSAT mission, in far UV by the IUE satellite, in visible by the Galileo spacecraft's SSI camera and
from the ground at infrared and radio wavelengths \citep{b1}.  In the powerful magnetosphere of Jupiter the
accelerating electrons and ions spiral down towards the polar regions while colliding with particles of
the planetary atmosphere. The auroral display in the atmosphere of Jupiter and other giant planets arises due to excitation
of hydrogen species rather than oxygen and nitrogen that give the characteristic colours of the aurora in the Earth's atmosphere.
\color{black} 
While the solar wind is thought to affect auroral emission on Jupiter the interaction of the moons, Io and
Ganymede with Jupiter's magnetosphere contributes most ionised particles that form the magnetised plasma torus around the planet.


The observed auroral emissions are dominated by excitation, ionisation and dissociation processes that are due to energetic electrons impacting on hydrogen molecules. Heavier ions  have also been  observed  in Jupiter's magnetosphere and are thought to
contribute to the aurora. Volcanic activity on Io was proposed as the origin of these particles that are swept up and ionised
in the planetary magnetosphere \citep{b3}. The de-excitation of hydrogen and its molecular form, H$_{2}$ leads to the UV emission that appears to have a persistent presence in the polar regions of the planet \citep{b60}. Its most stable component is a characteristic, intense main oval emission, which forms due to the co-rotation breakdown of the plasma escaping radially from the torus. The Alfvenic currents generated at Io and other Galilean moons form arcs with bright spots of auroral emission in the vicinity of the main oval. These, so called footprints, change longitudinal position due to orbital motion of the moons within the magnetosphere co-rotating with the planet \citep{b2}. The irregular in shape, highly variable, polar emissions are also visible inside the main oval. 

On the other hand the ionized H$_{2}$ combines with a molecular neutral hydrogen in exothermic reactions leading to formation of the H$_{3}^{+}$ molecule responsible for auroral emission in the infrared.
\color{black}
Observations of Jupiter in the far infrared also revealed strong limb brightening at high latitudes that was identified
as emission due to bands of methane at 7.7~$\mu$m, acetylene at 13.6~$\mu$m, ethane at 12.2~$\mu$m and other hydrocarbons
\citep{b4}.

The H$_{3}^{+}$ molecule was first identified in laboratory plasmas produced in discharge cells from H$_{2}$ gas \citep{b7}.
It is the most abundant product of hot environments dominated by hydrogen. Therefore it is important in astrophysical
plasmas such as planetary ionospheres and parts of the interstellar medium as a cooling agent during stellar formation
\citep{b8}. H$_{3}^{+}$ is the simplest highly symmetric, triatomic molecule that makes theoretical {\it ab initio}
calculations feasible. Such models show a good agreement with experimental spectra from laboratories and astronomical
objects. The rotational spectrum of the molecule arises due to centrifugal distortion that leads to a small dipole moment.
Such a ``forbidden" spectrum has reasonably strong transitions due to the small nuclear mass of H$_{3}^{+}$. Its
vibrational spectrum with two modes $\nu_{1}$ and $\nu_{2}$, leads to a complex structure of transitions in fundamental,
overtone, forbidden and combination bands. The fundamental band, $\nu_{2} \rightarrow 0$, located in the infrared region
centred on 2521~cm$^{-1}$ produces the strongest transitions. Although typical overtone transitions tend to be very weak in
comparison, the H$_{3}^{+}$ molecule has much stronger first overtone $2\nu_{2} \rightarrow 0$, transitions due to small
masses and its largely anharmonic potential \citep{b9}.  

Both fundamental and overtone bands have been observed in Jupiter's ionosphere. Many auroral emission features of  H$_{3}^{+}$ were first detected in
low-resolution spectra at 2~$\mu$m by \citet{b5} who were only able to identify H$_{2}$ S(1) and S(0) quadrupole lines. An
assignment of other features followed from the work of \citet{b6}, who observed Jupiter with the Fourier Transform
Spectrometer (FTS) at  the Canada-France-Hawaii Telescope between 2.0 and 2.2~$\mu$m, and assigned 23 lines to the
H$_{3}^{+}$ molecule in the 2$\nu_{2}$ overtone band. This discovery prompted observations of the predicted to be stronger, 
fundamental $\nu_{2}$ band of H$_{3}^{+}$ at 3.5$-$4 $\mu$m \citep{b10,b11,b12}. 
Transitions observed in both bands are useful probes of the physical conditions in the upper atmosphere of Jupiter.
Measurements of temperatures, column densities and ortho-to-para ratios were followed by temporal monitoring and spatially
resolved mapping of auroral regions \citep{b13,b14,b15}.

In this paper we present GNIRS/Gemini spectroscopic observations from the Jovian auroral regions (Section~\ref{sec2}). We
describe the application of the ATMOF modelling technique that allows effective removal of telluric absorption that is important in detecting
weak spectral features (Section~\ref{sec3}). In Section~~\ref{sec4} we identify emissions from H$_{3}^{+}$ and H$_{2}$ and
propose assignments for the remaining lines.  Next we determine the rotational temperature, column density and vertical
profiles of H$_{3}^{+}$ in the ionospheric region of the planetary atmosphere together wih estimates of its maximum
density (Section~\ref{sec5} and~\ref{sec6a}). We also use H$_{2}$ quadrupole lines detected in the low-resolution spectrum of the northern
auroral region to derive rotational temperature for this molecule (Section~\ref{sec6}). Finally we discuss and compare our
measurements with previously published observations in Sections~\ref{sec7} and~\ref{sec8}.

\section{Observations and Data }
\label{sec2}

We carried out high-resolution, long slit spectroscopy of the Jovian central meridian in the K band between 2.01 and 2.07~$\mu$m
with the GNIRS instrument on the GEMINI North 8m telescope in August, 2011 (Table~\ref{tab:1}) as a part of the project to understand
the D/H ratio in giant planets and Titan \citep{b36,b20}. 

\begin{figure}
  \centering
 \includegraphics[width=8cm,angle=270]{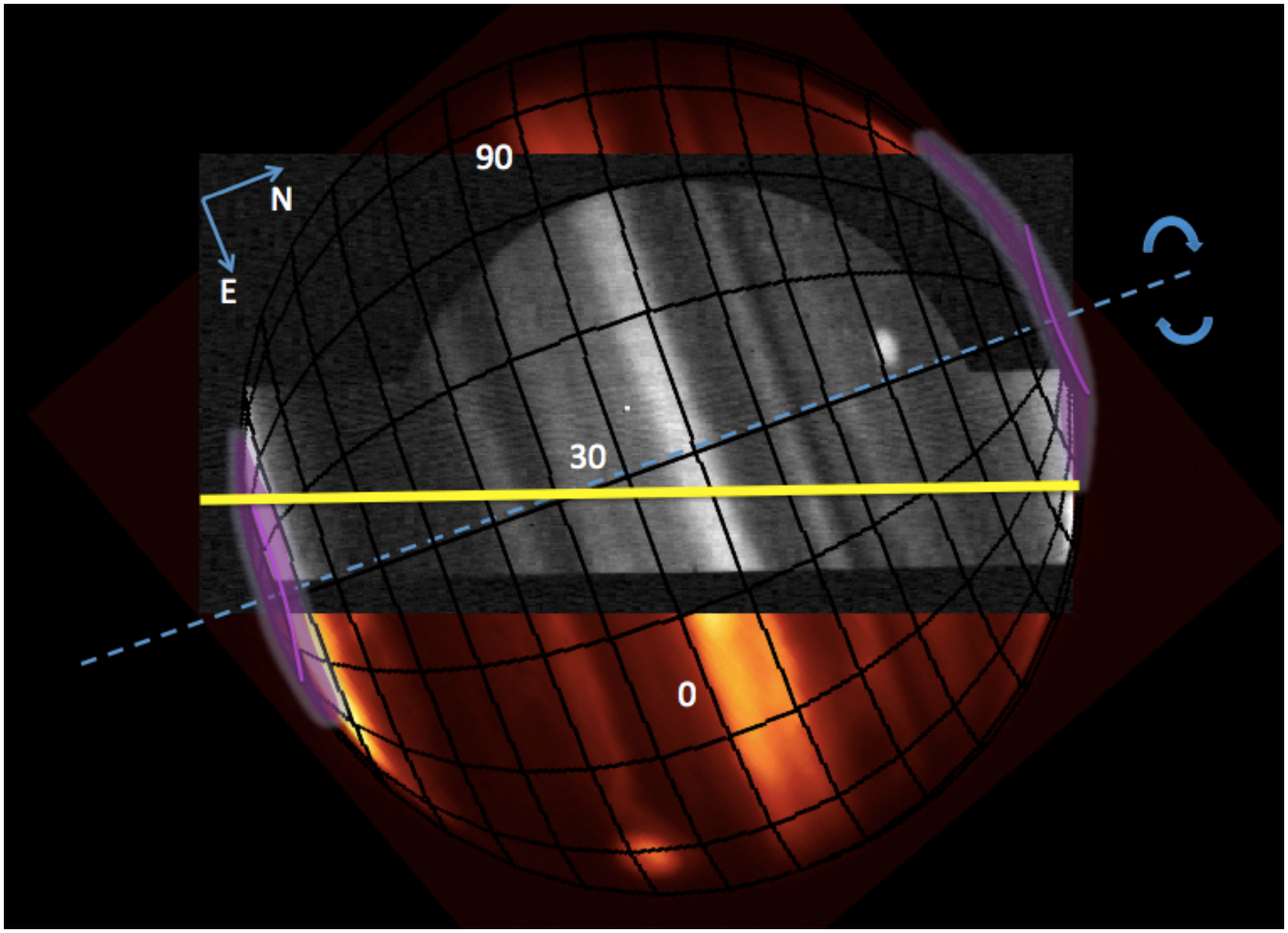}
  \caption{
 \color{black}
Acquisition image of Jupiter in the narrowband H$_{2}$ filter at 2.04~$\mu$m (in black-and-white) overlaid on the 2.1~$\mu$m image of the planet taken with the IRTF on July 19, 2009 (Credit: NASA/IRTF/JPL/University of Oxford). At the time of our observations the North pole was 20.6 arcsecond from the sub-observer point. The apparent longitude of the disk centre is about 31$^{o}$  in the $\lambda_{III}$ coordinates system. The slit position is marked with the yellow line. The approximate extent of the UV auroral regions including the Io footprint, are plotted in purple, with purple lines denoting approximate position of the main ovals. The polar projections of these regions is shown in more detail in Figure~\ref{fig:1a}. 
\color{black}
}
  \label{fig:1}
\end{figure}

The 0.1-arcsecond-wide slit positioned across the diameter of the planet in a direction encompassing the northern and
southern polar regions, is marked as the yellow line in Figure~\ref{fig:1}. The slit was tilted at close to 22 degrees with respect to the
planet's rotation axis, which was a requirement of the main project but it proved less convenient for observations of
auroral emissions. 
\color{black} 
In the northern polar region (Figure~\ref{fig:1} and~\ref{fig:1a}) the slit was only grazing the outer edges of the typical extent of the UV aurora encompassing the Io footprint as defined in \citet{b60}. In Figure~\ref{fig:1a} the expected position of the main Io footprint spot \citep{b77} is marked with the square on both northern and southern aurora. The size of both squares corresponds to measurement errors given in that paper. Interestingly, our northern spectrum falls onto the region, which encompasses the expected auroral emission from the main Io spot. In the south the slit cuts through the inner aurora, which should include the emissions associated with the main oval. 

\begin{table*}
\footnotesize
 \centering
 
\begin{minipage}{20cm}
 \caption{Observing conditions and settings of the GNIRS instrument} 
 \begin{tabular}{ccccccc}
    \hline \hline

 &\textbf{Jupiter} &\textbf{Central }	&\textbf{} & \textbf{Integration} &\textbf{Mean} & \\
  \textbf{Date UT }&\textbf{CML III}&\textbf{wavelength}	&\textbf{Resolution} & \textbf{ time} &\textbf{ airmass} & \textbf{Seeing}\\
	& {\it deg}   &{\it $\mu$m} &	 & {\it sec} &  & {\it arcsec}\\
\hline
24/08/2011 15:09  & 300 &   2.08 &      5900 &   60 &1.01&0.4 \\
27/08/2011 15:01  &  31  &   2.04& 18000 &160 & 1.009 & 0.25\\

 \hline
 \end{tabular} 
 \label{tab:1}
\end{minipage}
\end{table*}
\color{black}
We used the 111 lines/mm grating to obtain a spectral resolution of R~$\sim $~18000.
These observations were complemented with lower-resolution spectra taken with the 32 lines/mm grating (R~$\sim $~5900)
to cover a broader wavelength region in the K band centred on 2.08~$\mu$m between 1.96 and 2.17~$\mu$m. We applied standard
calibration techniques for the 2-D spectral images by using the Gemini IRAF GNIRS package. The images were aligned and the
instrument specific noise due to variable bias was removed. Next the non-linearity correction was performed and the bad
pixels on the detector frames were flagged. After that, images were flat-fielded and wavelength calibrated with Ar/Xe lamps. 
Finally the frames were sky-subtracted and the s-distortion correction was applied. 

\begin{figure*}
\centering
\includegraphics[angle=270,trim={8cm 5.5cm 7cm 5cm},clip,width=14cm]{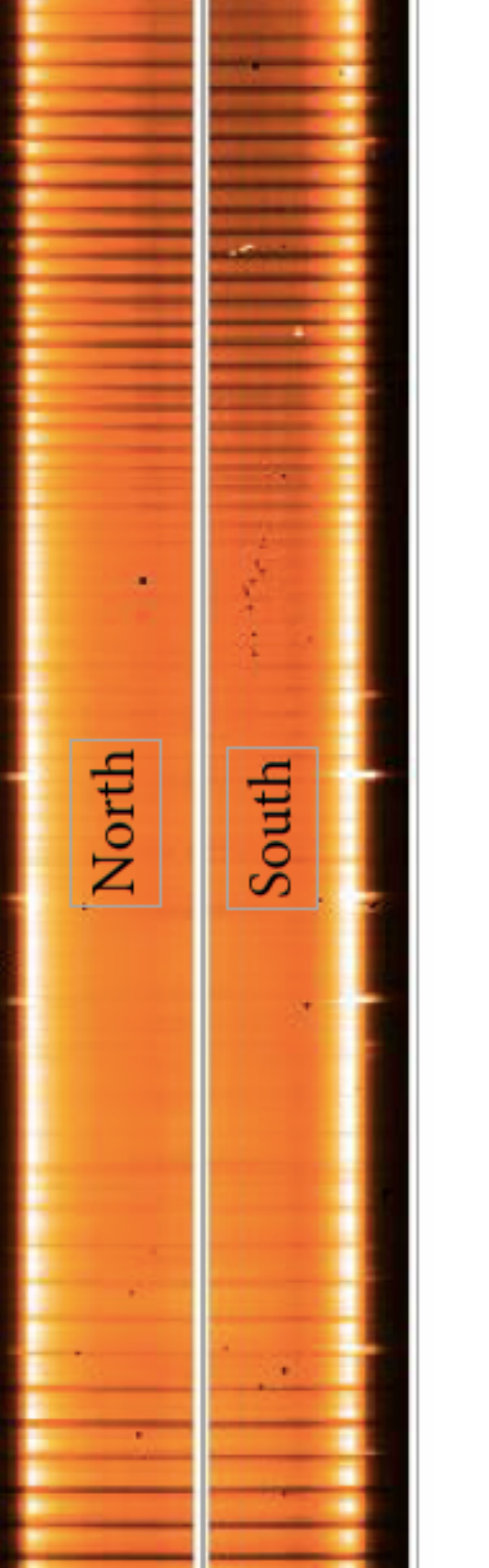}
 \caption{
2-D raw spectra of both polar regions across 60 pixels show auroral emissions at the edge of the planetary limb that are distinct from the brightening of the polar continuum in K band between 2.01 and 2.07~$\mu$m.  Spectral resolution is R~$\sim$~18000. 
\color{black}
}
\label{fig:2}
\end{figure*}

\begin{figure}
  \centering
 \includegraphics[width=11cm]{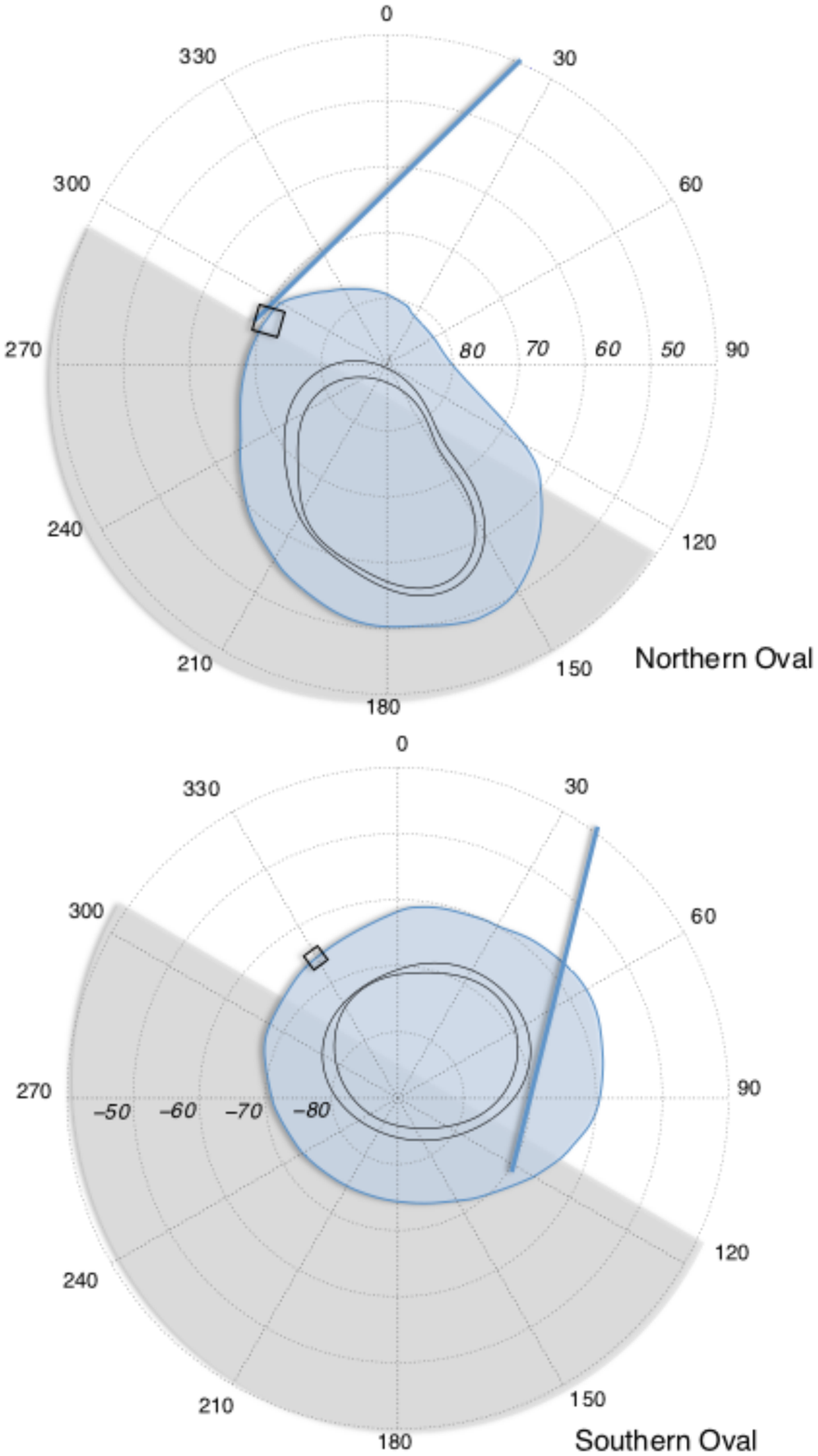}
  \caption{
The northern and southern polar projections showing the extent of the typical UV aurora oval and Io footprint as defined in \citet{b60}. The shaded regions define the night time during our observations. The slit position is approximated with the blue, straight lines. The black squares denote the expected position of Io footprint during our observations according to data from \citet{b77}. 
 \color{black}
}
  \label{fig:1a}
\end{figure}

The auroral emission lines were clearly visible in the 2-D high-resolution spectra in both polar regions 
(see
Figure~\ref{fig:2} that shows the northern and southern spectral region with strong emission lines from H$_{3}^{+}$ at the edge of the
limb of the planet). 
\color{black}
We extracted 1-D spectra for all spatial rows with visible auroral emissions.
The pixel size of 0.05$^{\prime\prime}$ for  Jupiter's diameter of 42.1$^{\prime\prime}$ at the time of observations
corresponds to about 167 km. In the southern region the peak of the auroral emission lines is offset by 6 pixels ($\sim$
1000 km) with respect to the peak of the CH$_{4}$ polar brightening \citep{b54} presumably due to the lower level
H$_{2}$-H$_{2}$ collisional absorption \citep{b23}.  In the northern region the corresponding offset is just 1 pixel from
the brightened edge of the limb. 
The northern pole was oriented towards the observer at a distance of 20.6$^{\prime\prime}$
from the centre of Jupiter's disc. In such a planetary orientation, taking into account the slit positioning
 in Figure~\ref{fig:1}, it is possible to see higher altitudes outside of the planetary limb in the southern polar region 
of the planet. The northern polar emissions can be visible above but mostly overlapping  the northern edge of the limb
in projection. 
\color{black}
Consequently the actual physical displacement of the aurora emissions from the brightened limb derived in
previous modelling \citep{b31} and observations \citep{b26} translates more easily to our estimates for the southern
region. 

The extracted 1-D spectra were affected by absorption features due to strong telluric bands of CO$_{2}$ in parts of the K-band
(regularly spaced dark lines in Figure~\ref{fig:2}). The usual procedure to deal with this contamination would be to divide
our target spectrum by the spectrum of a telluric standard star that has been observed in temporal and spatial proximity to
the target object. This would serve to remove any solar and telluric features resulting from the Earth's atmosphere.  Here
we applied a novel, more sophisticated and precise technique of telluric subtraction \citep{b16} that is described in the
next section.  The extracted spectra of the northern and southern aurora were, in the first instance averaged over 4 pixels around the peak
emissions of H$_{3}^{+}$. Due to non-symmetric positioning of the long slit and its smaller extent for the low-resolution
settings, such spectra extracted from the northern regions are closer to the edge of the planetary limb. The most affected is
the low-resolution spectrum of the northern aurora, where the emission features show only weakly against the stronger
continuum from the disk of the planet.

\section[]{ATMOF Telluric Model} \label{sec3} The removal of the telluric absorption from our spectra was carried out using
the ATMOF (ATMOspheric Fitting) code and related procedures  \citep{b16}.  This method was previously applied successfully
to remove telluric lines from spectra of Jupiter and Titan, as well as Neptune \citep{b20}. A spectrum with telluric
features that correspond to our observations is derived by a modelling process, where a solar spectrum is passed through the
Earth atmosphere model and the telescope instrumental response is subsequently applied. A number of free parameters such as
CO$_{2}$, H$_{2}$O content and the Doppler shift of the standard star are fitted to match the model spectrum to the standard
star (HIP11503) data. 

\begin{figure}
 \includegraphics[width=8.5cm]{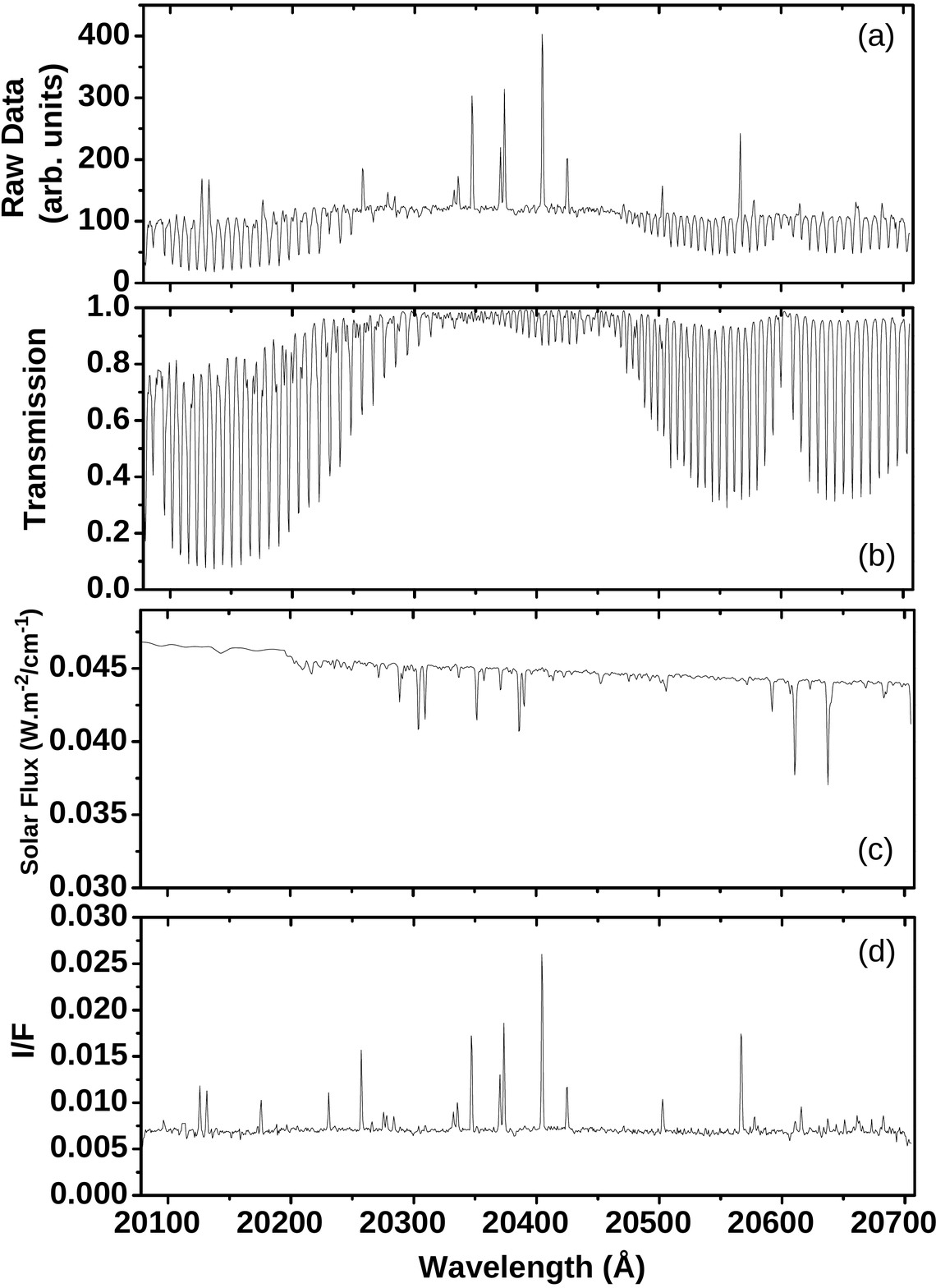}
 \caption{The spectral components used in the ATMOF modelling to remove telluric and solar features from the high-resolution spectrum of the southern auroral emission. (a) The 1-D spectrum of auroral emission before telluric subtraction, (b) The atmospheric model that provided best fit to the telluric features of CO$_{2}$ and H$_{2}$O absorption matching the standard star spectrum as applied to the Jovian spectra at the zenith angle of Jupiter's observations,  (c) Solar spectrum described in Section~\ref{sec3}, (d) Final 1-D spectrum after application of the ATMOF model. }
\label{fig:3}
\end{figure}

We used VSTAR - Versatile Software for the Transfer of Atmospheric Radiation \citep{b17} to obtain the Earth atmosphere
models for Mauna Kea Observatory with varied atmospheric content of carbon dioxide and water vapour. The CO$_{2}$ mixing
ratio is varied as a fraction of the whole altitudinal profile, whilst the H$_{2}$O vapour mixing ratio is adjusted
gradually only for the lower layers of the atmosphere as detailed in \citet{b16}. In Figure~\ref{fig:3}(b) the model is shown
that best matches telluric features present in the standard star.  The telescope instrumental response consists of a scaling
factor and slope, a wavelength shift correction, and the filter response for the GNIRS K-band order-blocking filter used in
our observations, which is largely flat in the spectral region under consideration.  In Figure~\ref{fig:3}(c) we show the
solar spectrum as a composite of high-resolution data (R = 300,000)  from the Kitt Peak Solar Atlas \citep{b19}, stitched
together by using the low-resolution synthetic solar spectrum derived by Robert Kurucz\footnote{Kurucz, R. L. Accessed
online on 12th January 2009. The solar irradiance by computation (http://kurucz.harvard.edu/papers/irradiance/solarirr.tab)}
as a template for the intensity scale. At wavelengths below 2.02 $\mu$m the Kitt Peak data is incomplete, therefore the
Kurucz's low-resolution synthetic spectrum is substituted in this region (see Figure 1 in \citet{b16}).  

Once the parameters for the Earth atmosphere model have been retrieved, the model is re-run with the zenith angle
corresponding to the Jupiter observations. The Jovian auroral spectra are then divided by the telluric transmission, and the
instrumental response with the retrieved parameters applied. This is followed by application of the wavelength shift to the
Jovian auroral data, which is determined independently for each observation by fitting a Doppler-shifted solar spectrum
passed through the Earth atmosphere model with telluric features. In this way the solar and telluric lines in the Jovian
spectra can be used simultaneously to fine-tune the wavelength shift. 

Finally we apply a baseline correction as a quadratic fit to the continuum that takes care of the residual background level
remaining after sky subtraction due to unaccounted scattered light from the planet. The baseline subtracted varied slightly
for all spectra, and was between 3 to 6.5\% of the maximum signal intensity. This procedure is applied to all spectra
extracted from the two-dimensional image shown in Figure~\ref{fig:2} for the range of pixels where northern and southern
auroral emission is visible. Also the 4-pixel-average of low and high-resolution spectra in Figure~\ref{fig:4} and
Figure~\ref{fig:5} have been separately corrected with the ATMOF technique.

\section{Identification of spectral lines in northern and southern regions}
\label{sec4}
We used the Specview\footnote{Specview can be found on the STScI website: 
 \newline http://www.stsci.edu/institute/software\_hardware/specview} tool for the analysis of 1-D spectra to fit Gaussian components to visible emission lines in order to derive their wavelengths and relative intensities. 
We fitted 26 emission lines in our high-resolution spectra. Additional emission features were fitted in the low-resolution datasets in spectral regions extending beyond the range of the high-resolution spectrum. Our detections of the H$_{3}^{+}$ transitions are listed in Table~\ref{tab:12}. 
In the high-resolution spectra we identified all the H$_{3}^{+}$ transitions expected in the region between 2.01 and 2.07 $\mu$m in temperatures T$_{rot}$~=~T$_{vib}$~=~1000~K from the database in \citet{b21}. In the low-resolution spectra only some of the brightest transitions were clearly visible (see Comments in Table~\ref{tab:12}). The strongest H$_{3}^{+}$  transition visible in our low-resolution spectrum is tR(6,6) at 2.093 $\mu$m - outside the range of the high-resolution spectra. 

\begin{figure*}
\includegraphics[width=14cm,angle=0]{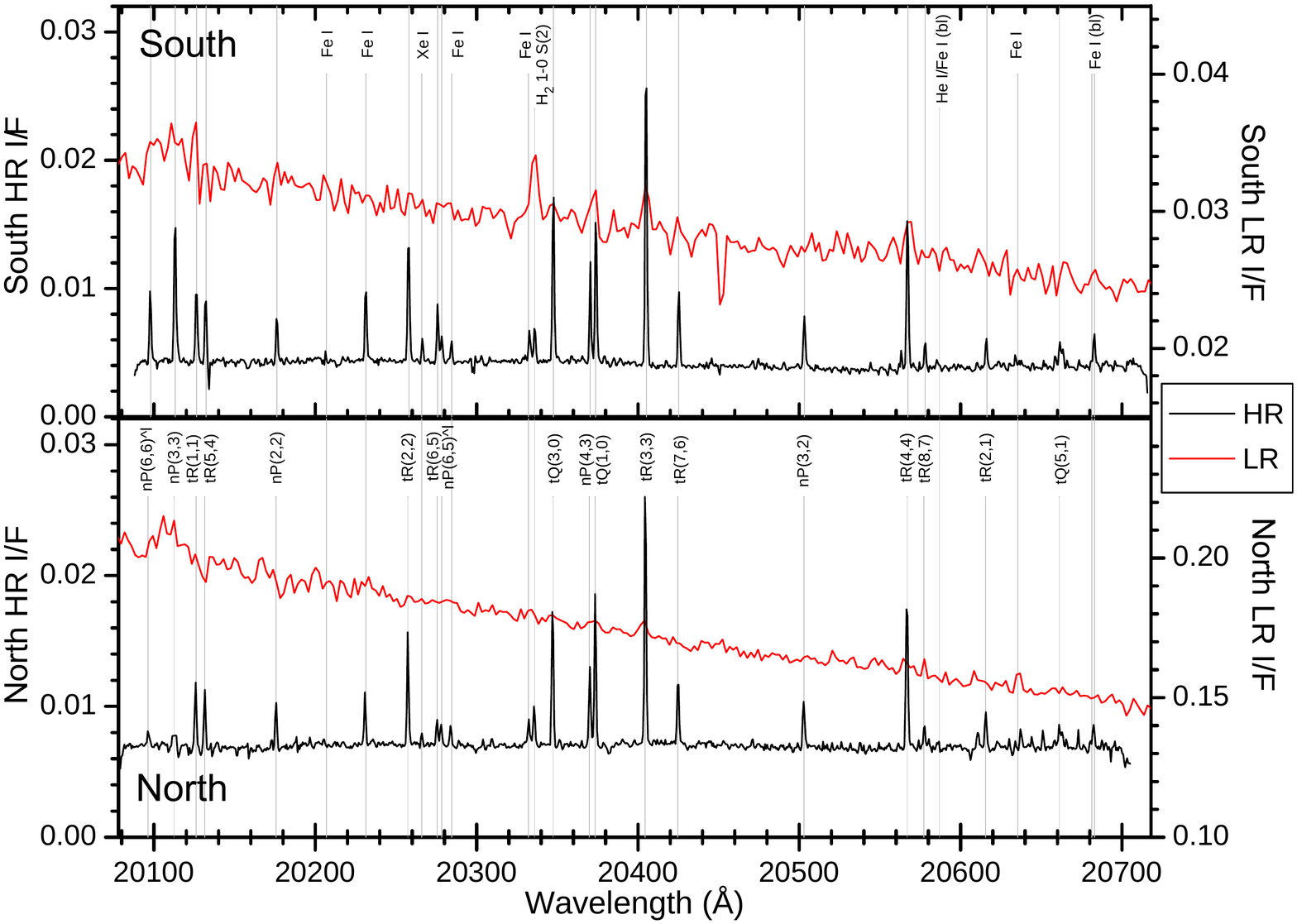} 
\caption{High-resolution (black) and over plotted low-resolution spectra in
red for the southern (upper panel) and northern (bottom panel) aurora are shown. The radiance factor, I/F where I is the reflected intensity of the planet and $\pi$F is the solar flux incident on the planet, as defined in
\citet{b23}. In the lower panel we marked the assignments of the H$_{3}^{+}$ transitions. In the upper panel the remaining
identifications together with the proposed Fe~I assignments are denoted.  } 
\label{fig:4} 
\end{figure*}

\begin{figure*}
 \includegraphics[width=14cm]{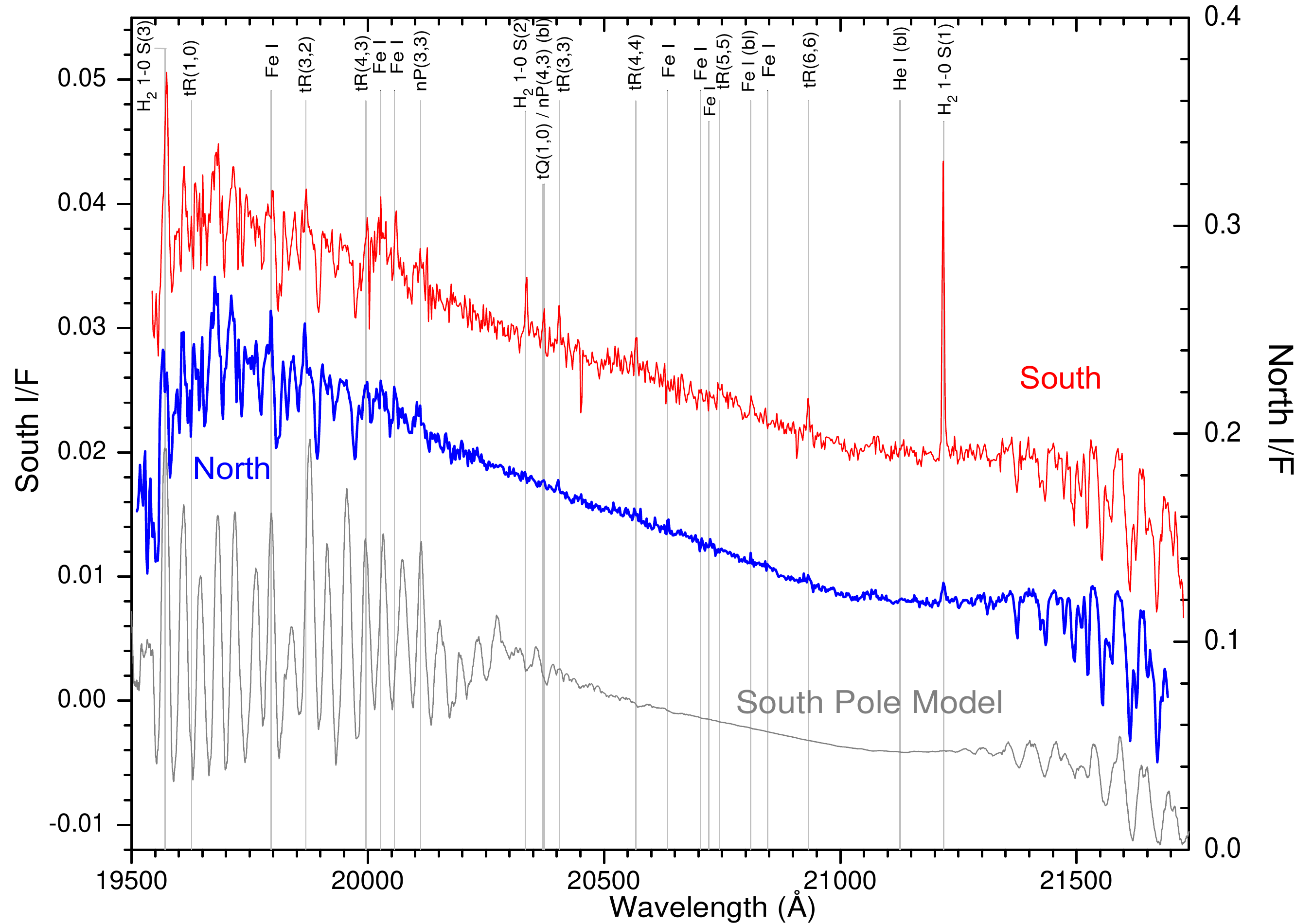}
 \caption{Low-resolution spectra for the southern (red) and northern (blue) aurora are presented. We also show a model of the atmosphere at the south polar limb of Jupiter, which was derived with the VSTAR modelling code. In the left and right ends of this spectral region there are strong bands of methane absorption in the atmosphere that make identification of only the strongest emissions possible when observed against the limb of the planet. We marked the positions of the lines expected in this part of the spectrum. The strongest lines belong to  H$_{2}$ 1-0 transitions. There are also strong H$_{3}^{+}$ transitions visible. We also marked positions of the strongest transitions of Fe~I that should be visible in the low-resolution spectrum.  }
\label{fig:5}
\end{figure*}

In Figure~\ref{fig:4} we marked the well known quadrupole transition from molecular hydrogen with H$_{2}$ 1-0 S(2) at
2.0338~$\mu$m in our high-resolution spectrum. In the low-resolution spectra for the southern region (Figure~\ref{fig:5})
the S(2) transition is also visible with the H$_{2}$ 1-0 S(1) and S(3) at 2.1218 and 1.9576~$\mu$m respectively. The
corresponding H$_{2}$ transitions in the northern region are less pronounced, because the slit position did not extend
beyond the planetary limb. 

The fitted wavelengths of the H$_{3}^{+}$ and H$_{2}$ lines in the high-resolution spectra show systematic shifts with
respect to the vacuum wavelengths. The average difference between laboratory and observational assignments for the northern
and southern aurora are 1.98 and 1.30~$\AA$  respectively. The relative average wavelength difference in H$_{3}^{+}$ and
H$_{2}$ identifications between the two auroral regions of 0.68~$\AA$  translates to 10.2 km/s difference in radial
velocity, which can be explained by different and opposite Doppler shifts due to the rotation of the planet. 

Beside the H$_{3}^{+}$ and H$_{2}$ dominant transitions there are many more emission lines that we attempted to identify
with different species that could be present in the Jovian aurora.  We detect the atomic He I emission at 2.0587 $\mu$m in the
high-resolution spectrum at low signal-to-noise level but with peak intensity that coincides with
neither, H$_{3}^{+}$ or H$_{2}$. Our high-resolution spectra contain many additional lines that have not been assigned
before. Some of these features could be consistent with the low intensity transitions of H$_{3}^{+}$, that are not present
in the \citet{b21} database. \citet{b26} list additional transitions that coincide with lines visible in our spectrum such
as P(5,4) at 2.0334~$\mu$m and P(5,3) at 1.9593~$\mu$m. It is not clear if the intensities of these additional assignments
are consistent with the theoretical calculations \citep{b29}. 

We also considered the possibility that atomic transitions could be present in the spectral range under consideration.  We
used the NIST database \citep{b22} to attempt matching any unassigned lines. Out of a large number of neutral and ionised
atomic transitions expected in this spectral region, a significant fraction of them belongs to different excitation levels
of Fe~I; these were found to match closely many weaker lines observed in our spectra (Table~\ref{tab:2}).  

\begin{landscape}
\begin{figure*}
 \includegraphics[width=17cm]{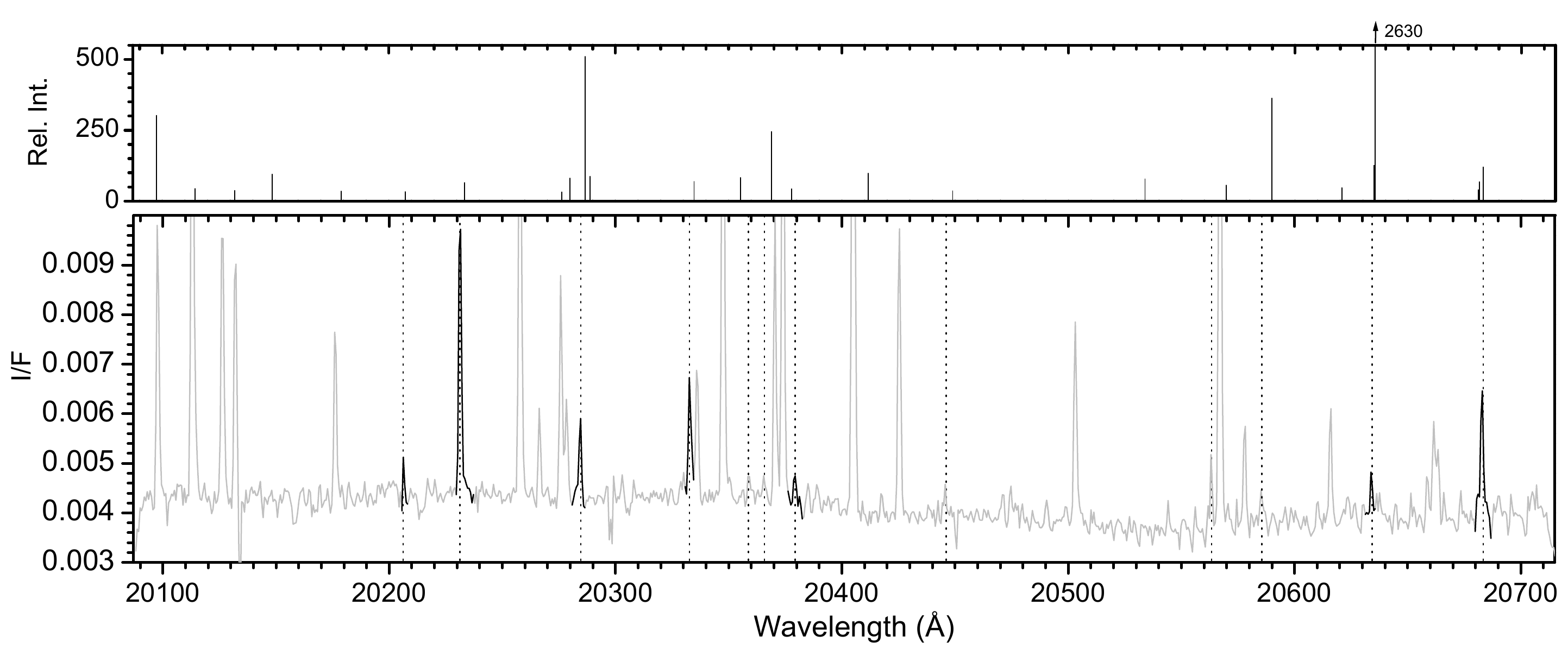}
 \caption{The Fe~I transitions listed in column 2 of Table~\ref{tab:2} are shown as a stick spectrum in the upper panel. Below we zoomed-in the high-resolution spectrum of the southern aurora to show the weak lines. It can be seen that the noise level increases only slightly at the edges of the spectrum due to removal of the strong CO$_{2}$ absorption bands. We marked the strong lines that we associated with the Fe~I  transitions in bold. The dotted lines are drawn to show the position of the closest unidentified lines observed near the listed Fe~I transitions. }
\label{fig:6}
\end{figure*}
\end{landscape}

We examined these assumed identifications and
fitted the strongest lines with the Specview tool. Their wavelengths and relative intensities are given in Table~\ref{tab:2}
for spectra from Figure~\ref{fig:4} and Figure~\ref{fig:5}. In Figure~\ref{fig:6} we zoomed in on the faintest transitions
in the high-resolution spectrum of the southern aurora and marked the lines coinciding with the strongest Fe~I lines
expected in this region. The noise level in this spectrum is estimated to be 0.5$\times10^{-7}$ W m$^{-2}$ sr$^{-1}$. In
Table~\ref{tab:2} we also list some of the strongest Fe~I features that could possibly be detected in the low-resolution
spectrum, especially for the southern auroral region, which extends further beyond the limb of the planet than the northern
region. We suggest that a strong emission at 2.0268~$\mu$m (Figure~\ref{fig:4}) could be matched with the strongest Xe~I
transition expected in our observed region between 1.96 and 2.17~$\mu$m. These possible assignments are discussed further in
Section~\ref{sub8.4}.

\begin{landscape}
\begin{table*}
\footnotesize
 \centering
 
\begin{minipage}{20cm}
 \caption{The list of emission lines that were identified as H$_{3}^{+}$ transitions in spectra of both polar regions of Jupiter. All lines belong to 2$\nu_{2}$~$\rightarrow$~0  emission band. Fitted wavelengths and intensities are provided for all lines visible in the high-resolution spectra. Out of five additional transitions that are present in the low-resolution spectra, we were able to identify four of them in the southern auroral region and three in the northern region. In comments we note which transitions were visible in the low-resolution spectra, where (n) and (s) mean northern and southern aurora respectively. }   
 \begin{tabular}{crccccccl}
    \hline \hline
\multicolumn{4}{l}{}& \multicolumn{2}{c}{\textbf{Southern Aurora}}& \multicolumn{2}{c}{\textbf{Northern Aurora}}&\\
  \textbf{Transition}&\textbf{Intensity\footnote{This is the integrated absorption intensity derived from the formula (3) in \citet{b21} using the Online H$_{3}^{+}$ Intensity Calculator from the H$_{3}^{+}$ Resource Center  (http://h3plus.uiuc.edu/criteval/calc.shtml) under assumption of T$_{rot}$~=~T$_{vib}$~=~1000~K.  }}	&\textbf{Vacuum} & \textbf{Branch \&} &\textbf{Observed}&\textbf{Intensity}&\textbf{Observed}&\textbf{Intensity}& \textbf{Comments}\\
 \textbf{Wavenumber}&			&\textbf{wavelength}	&\textbf{Transition }  &\textbf{wavelength}	& \textbf{$\times10^{-7}$}&\textbf{wavelength}	& \textbf{$\times10^{-7}$}& \\
\hline
cm$^{-1}$	&   cm$^{-2}$ atm$^{-1}$&	\AA 	& & \AA & Wm$^{-2}$sr$^{-1}$&\AA&Wm$^{-2}$sr$^{-1}$&\\
\hline
4732.041   &      5.238062 & 21132.53&tR(7,7) & 21134.5& 1.86& 21101.4& & low(n.s)\\
4777.226  &     13.509024 &      20932.65 &        tR(6,6) & 20932.1&   4.60&20925.2&   &low(n,s) \\
\multicolumn{9}{l}{\it High-resolution region starts}\\
4839.508	 &	1.645477	 &	20663.26	&	  tQ(5,1)& 20663.45&    5.86 &20661.21 &   3.33& \\
4850.264	 &	2.021775	 &	20617.43	&	  nP(2,1)& 20616.26 &   3.00 &20615.41&   4.56&\\
4859.212	&	1.676679	 &	20579.47	&	  tR(8,7)& 20578.04&     2.40&20577.32&   3.23& \\
4861.790  &	8.928174	 &	20568.56	&	  tR(4,4)& 20567.43&     14.93&20566.92&   22.63&low(n,s)\\
4876.938	 &	3.227214	 &	20504.67	&	  nP(3,2)& 20503.10&   4.82&20502.78&   6.91&\\
4895.518 	 &	5.027651	 &	20426.85	&	  tR(7,6)& 20425.37&     6.77&20424.77&   10.41&\\
4900.393  &    17.556369  &	20406.53 &	 tR(3,3)& 20405.19&     27.27&20404.42&      36.00&low(n,s)\\
4907.871	 &	9.974512	 &	20375.43 &	 tQ(1,0)& 20373.72&      12.20&20373.34&    18.93&low(s)\\
4908.672	 &	6.496212	 &	20372.11	&	  nP(4,3)& 20370.42&    6.77 &20369.91&    10.72&\\
4914.248 &     11.810715  &	20348.99 &         tQ(3,0)& 20347.58&     14.29&20347.42&   20.27&low(s)\\
4930.981	 &	1.736462	 &	20279.94 	&	  nP(6,5)$^{l}$& 20278.44&  1.99&20278.31& 3.01&\\
4931.596	 &	3.475673	 &	20277.41 	&	  tR(6,5)&20275.90&      4.55&20275.61&   3.71&\\
4936.000	 &	7.481089	 &	20259.32	&	  tR(2,2)&20258.11&      11.83&20257.39&   14.39&\\
4955.991	 &	4.101676	 &	20177.60	&	  nP(2,2)&20176.11&     3.03&20175.70&   5.71&\\
4966.838	 &	4.391920	 &	20133.53	&	  tR(5,4)&20132.39&      4.21&20131.44&   7.26&\\
4968.272	 &	5.010080	 &	20127.72	&	  tR(1,1)&20126.30 &     5.91&20126.04&   8.15&low(s)\\
4971.561	 &   10.582623   &	20114.41	&	  nP(3,3)&20113.18&     12.83&20112.45&   14.44&low(n,s)\\
4975.338	 &	4.023513	 &	20099.14	&	  nP(6,6)$^{l}$&20097.93&   4.93 & 20096.28&   8.26& \\
\multicolumn{9}{l}{\it High-resolution region finishes}\\
5000.499 & 9.807676 & 19998.00 & tR(4,3) & 19999.0 & 5.78 &  \multicolumn{2}{c}{on CH$_4$ absorption} &low(s) \\
5032.447 &      4.824500  &  19871.05 & tR(3,2) & 19869.8 & 7.81 &19870.0 & & low(n,s) \\
5094.218 &  8.831507 & 19630.10 & tR(1,0) &   \multicolumn{4}{c}{severely affected by CH$_4$ absorption} &  \\ 
 \hline
 \end{tabular} 
 \label{tab:12}
\end{minipage}
\end{table*}
\end{landscape} 

\begin{landscape}
\begin{table*}
\tiny
  \centering
\begin{minipage}{20cm}
 \caption{The strongest Fe~I transitions (above relative intensity 30) as listed in the NIST ASD database \citep{b22} are reproduced in columns 2 and 3.  The fitted corresponding wavelengths and intensities from the southern auroral region (unless indicated otherwise in Comments) are shown in column 4 and 5 respectively. These were derived predominantly from the high-resolution spectrum of the southern aurora. We show fits for only these lines that have a peak intensity above 1$\sigma$ noise level estimated from the continuum between the strong lines. The low(n), low(s) and low(n,s) in Comments column denote possible detections in the low-resolution spectra either in northern, southern or both auroras respectively. For lines identified from the low-resolution spectra, only the wavelength of the peak emission is listed, because we found the intensity estimates difficult due to overlap with the strong methane absorption.}   
 \begin{tabular}{llrlrl}
    \hline \hline
  \textbf{Transition}&\textbf{ Observed}	&\textbf{NIST ASD} & \textbf{Fitted}& \textbf{Intensity}& \textbf{Comments}  \\
 \textbf{Wavenumber}&\textbf{Wavelength}&\textbf{Intensity \footnote{The relative intensity as shown in the NIST ASD database \citet{b22}, that use qualitative estimates taken from the sources referenced there.}} &\textbf{Wavelength} &   \textbf{$\times10^{-7}$} &      \\
\hline
cm$^{-1}$	&   	\AA& 	& \AA & W m$^{-2}$ sr$^{-1}$ &\\
\hline

5052.581202 & 19791.864 & 3160 & 19798.514 & & low(s) \\
4992.687211 & 20029.294 & 1480 & 20025.744 & & low(n,s) \\
4984.629397& 20061.672 & 33 & 20060.391 && low(n,s), inconsistent intensity, misidentification?  \\
\multicolumn{6}{c} {}\\
\multicolumn{6}{l}{\it High-resolution region starts}\\
4975.732358 & 20097.544 & 302 &  & & overlap with nP(6,6)$^{l}$ \\
4971.542889 & 20114.48 & 44 & & &  overlap with nP(3,3) \\
4967.241045 & 20131.9 & 37 &  & &  overlap with tR(5,4) \\
4963.134334 & 20148.558 & 95 & & & possible feature within 1$\sigma$ noise level\\
4955.666607 & 20178.92 & 35 &  & & overlap with nP(2,2) \\
4948.723309 & \textbf{20207.232} & 33 & 20206.111 & 1.09&  \\
4942.298419 & \textbf{20233.501} & 65 & 20231.299 & 6.60& \\
4931.858971 & 20276.33 & 32 & & & overlap with tQ(6,5) \\
4930.995404 & 20279.881 & 81 &  & & overlap with nP(6,5)$^{l}$ \\
4929.357379 & \textbf{20286.62} & 510 & 20284.676 & 1.99 & \\
4928.814366 & 20288.855 & 87 &(20289.693)\footnote{The wavelength estimate was derived from the peak in the northern spectrum.}&& marginal detection only in the northern spectrum\\
4917.691370 & \textbf{20334.745} & 69 & 20332.697 & 3.05 & blends with H$_2$ S(2) in low res. spectrum\\
4912.731225 & 20355.276 & 83 & 20358.754 & 0.59 & \\
4909.462154 & 20368.83 & 245bl & 20365.762 & 0.67& blends with nP(4,3) and tQ(1,0) in low res. spectrum\\
4907.298674 & \textbf{20377.81} & 43bl & 20379.371 &0.67 & \\
4899.185300 & 20411.557 & 98 &  & & no clear detection \\
4890.235007 & 20448.915 & 36bl & 20446.095 & 0.76 & \\
4870.025354 & 20533.774 & 78 &  & &  no emission\\
4861.549393 & 20569.574 & 56 & 20563.438 & 2.05& overlap with tR(4,4)\\
4856.771384 & 20589.81 & 363bl & 20585.602 &0.98 &blend with He I at 20586.92\AA~line  \\
4849.462970 & 20620.84 & 47 &  & & no emission \\
4846.158223 & \textbf{20634.902} & 126 & & & blend with 20635.327\AA~line \\
4846.058412 & \textbf{20635.327} & 2630 & 20634.322 &1.43 & low(n) \\
4835.373426 & 20680.926 & 40 &  & & blend with 20683.001\AA~line? \\
4835.276397 & \textbf{20681.341} & 68 &  & & blend with 20683.001\AA~line \\
4834.888322 & \textbf{20683.001} & 120 & 20683.451 & 3.58& \\
\multicolumn{6}{l}{\it High-resolution region finishes}\\
\multicolumn{6}{c} {}\\
4829.993409 & 20703.962 & 600 & 20703.964 & & low(n,s) \\
4825.647428 & 20722.608 & 600 & 20722.597 & & low(n,s) \\
4806.465273 & 20805.310 & 500 &20811.383  & & low(n,s) blend with 20810.774\AA~line  \\
4805.203305 & 20810.774 & 690 & 20811.383 & & low(n,s) blend with 20805.310\AA~line \\
\hline
 \end{tabular} 

\label{tab:2}
\end{minipage}
\end{table*}
\end{landscape}

\section{Determination of average temperature and column density}
\label{sec5}

Infrared auroral emission spectra of the H$_{3}^{+}$ have been used before to estimate the temperature of the upper
thermosphere of the planet. Most of the early research was focused on the $\nu_{2}\rightarrow0$, fundamental emission
spectrum at 3 to 4 $\mu$m that is much stronger than the $2\nu_{2}\rightarrow0$, overtone spectrum, with formally
``forbidden" transitions  \citep{b6,b12,b25,b13}. More recently observations of the H$_{3}^{+}$, overtone spectra in
selected regions of the K-band were investigated by independent groups \citep{b15,b26}. In 1999 and 2000 \citet{b15} imaged
the auroral emission of H$_{3}^{+}$, calculated its rotational, vibrational temperatures (T$_{rot}$,  T$_{vib}$) and column
densities, N(H$_{3}^{+}$) by using the high-resolution spectral region between 4700 and 4800~cm$^{-1}$ (2.08 - 2.12~$\mu$m)
with 13 overtone and 2 combination transitions. 

In our analysis we compare two ways of deriving rotational temperature for our set of the H$_{3}^{+}$ identifications in
averaged spectra at the peak emissions as defined previously (see Figure~\ref{fig:4} and Figure~\ref{fig:5}). First we use
the same method as \citet{b15} under the assumption that emission regions are optically thin and local thermodynamic
equilibrium (LTE) can be applied.  We note that  the ratios of all H$_{3}^{+}$ lines with respect to one specific line, can
be fitted to obtain the T$_{rot}$  by use of the following relationship:

\begin{equation} 
\label{eq:1}
T_{rot}=\frac{E'_{i}-E'_{0}}{k}\left[ln\frac{g_{i}(2J'_{i}+1)\nu_{i}A_{i}}{g_{0}(2J'_{0}+1)\nu_{0}A_{0}}-ln\frac{I_{i}(\nu_{i})}{I_{0}(\nu_{0})}\right]^{-1}
\end{equation}

where $J^{\prime}$ is the rotational quantum number of the upper level, while $E^{\prime}$ is its corresponding energy. The
$\nu$ is the transition frequency in cm$^{-1}$, $A$ is the Einstein A coefficient for a given transition and $k$ is the
Boltzmann constant. The nuclear degeneracy factor, $g$ is either 2, 4 or 8/3 (the modelling and assignment of these factors
is explained further in \citet{b21}). The zero subscript denotes the parameters for the reference transition, which was
chosen to be nP(3,2) at 2.0504 $\mu$m. This line was reasonably strong and well separated from any neighbouring emissions
that could affect its intensity. Here we make an important simplifying assumption that auroral regions can be described by
the single kinetic temperature equal to rotational temperature derived from the H$_{3}^{+}$ emission (T$_{kin}$=T$_{rot}$). 

To derive the fit for the temperature of both the northern and southern aurora regions, we used 16 lines from the high
resolution averaged spectra of H$_{3}^{+}$ with the exception of the tQ(5,1) and tR(6,5) transitions, which have
significantly different intensities with respect to other lines in both spectra; they are clear outliers and blending with
other nearby lines is a possible reason for inconsistencies in their intensity ratios. Our derived rotational temperatures,
T$_{rot}$ for northern and southern auroral region are 947~$\pm$~43~K and 916~$\pm$~58~K respectively.

In the second method for determining the rotational temperature, we simultaneously fit for rotational temperatures,
T$_{rot}$, and column densities along the line of sight, N(H$_{3}^{+}$). This is done with the assumption that, for
optically thin emission in LTE, the column density of the H$_{3}^{+}$ emitting plasma is related to the intensity of lines
and temperature as follows: 
\begin{equation} 
\label{eq:2}
I_{i}(\nu_{i})=\frac{N(H_{3}^{+})g_{i}(2J'_{i}+1)hc\nu_{i}A_{i}}{4\pi Q(T)} \exp^{\frac{-E'_{i}}{kT}}, 
\end{equation} where
{\it Q(T)} is a partition function at temperature T, while {\it c} is the speed of light. Specifically, this expression can
be rewritten as 
\begin{equation}  
\label{Eq:FitLine} y=C-\frac{x}{T}, 
\end{equation} with definitions 
\begin{gather}
\label{Eq:FitDefinitions} \nonumber C = \ln\left(\frac{N(H_{3}^{+})}{Q(T)}\right), \\ \nonumber x = \frac{E'_{i}}{k}, \\ y =
\ln\left(\frac{4\pi I_{i}(\nu_{i})}{g_{i}(2J'_{i}+1)hc\nu_{i}A_{i}}\right). \end{gather}

\begin{figure*}
 \includegraphics[width=10cm]{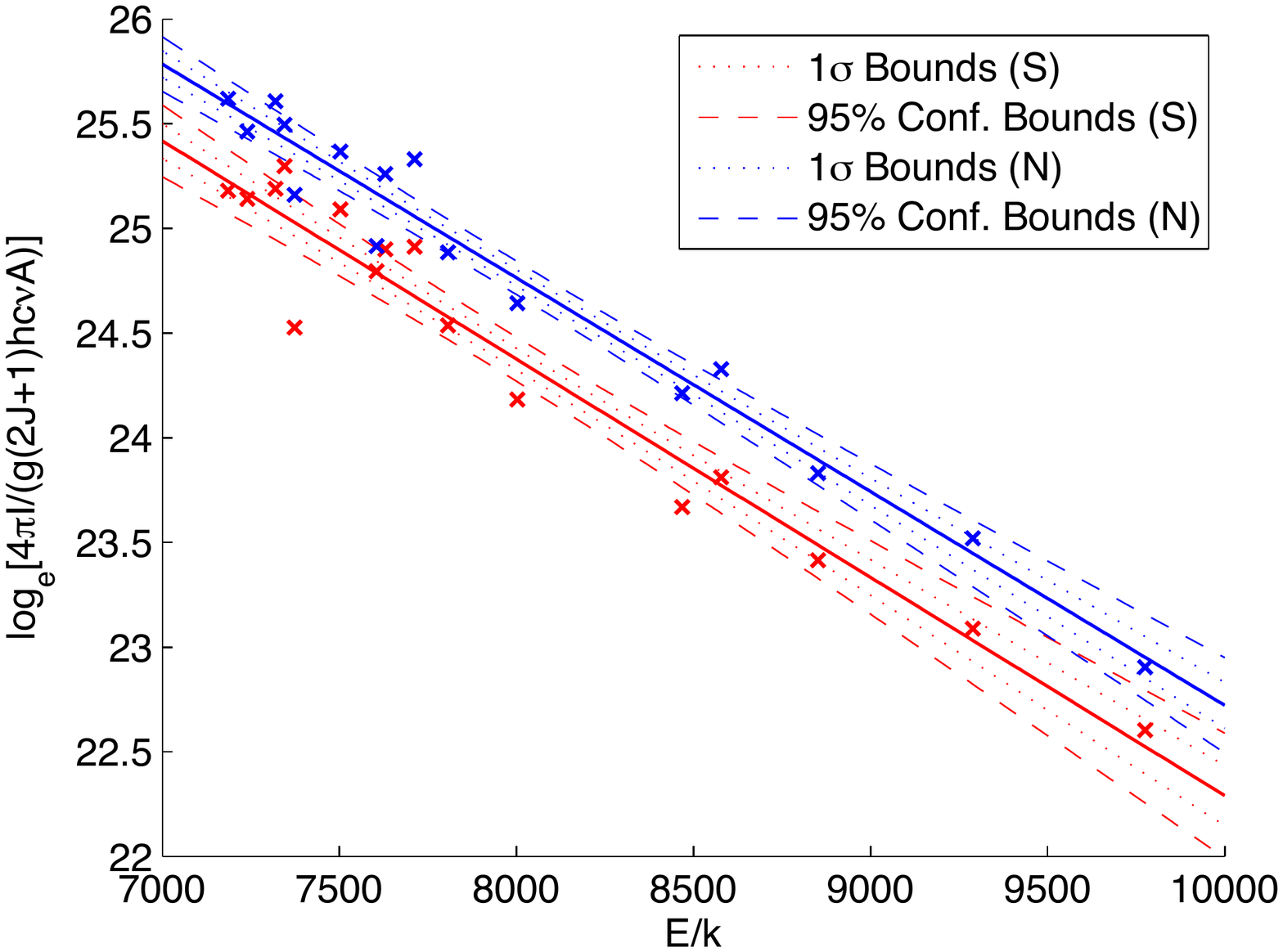}
  \caption{Lines of best fit for northern (blue) and southern (red) aurora data, according to Eq.~(\ref{Eq:FitLine}). The functionals are bound by the dotted curves when one standard deviation of both $C$ and $T$ parameters is taken into account. They are also bound by the dashed curves with $95\%$ confidence.}
\label{fig:7}
\end{figure*}

Accordingly, we numerically fit a line to the $(x,y)$ coordinates 
via the standard Levenberg-Marquardt algorithm, as implemented in MATLAB. 
\color{black}
 Confidence intervals for $C$ and $T$ can then be calculated using the inverse of Student's t cumulative
distribution function. Consequently, the linear fits for both the northern and southern aurora regions are shown in
Figure~\ref{fig:7}, as well as boundaries that account for a standard deviation in $C$ and $T$ values and boundaries that
represent a $95\%$ confidence interval for the fitting function.

From the slopes of these fits, we derive T$_{rot}$ temperatures of 979~$\pm$~49~K and 959~$\pm$~61~K  for the northern and
southern auroral regions, respectively. These values are slightly  higher than the temperatures obtained from the first method,
but are in agreement within their uncertainties.  Column densities can then be derived from the y-intercepts, $C$, once $T$
is known. So, for the northern aurora, possible densities range from 1.2$\times10^{16}$ to 7.0$\times10^{16}~m^{-2}$  with the best estimate as 3.1$\times10^{16}~m^{-2}$.  Likewise, for the southern aurora, the best estimate is
 5.2$\times10^{16}~m^{-2}$, with one standard deviation in functional parameters resulting in a possible density range from 2.6$\times10^{16}$ to 8.1$\times10^{16}~m^{-2}$.  Although both methods are equivalent, using only relative line intensities
in the first approach results in slightly reduced uncertainties on derived rotational temperatures. 
\color{black}Given that auroral components sampled in the north and south are due to interaction of different parts of the magnetosphere with Jupiter's ionosphere, the similar temperatures and column densities in both regions are noteworthy.\color{black}

\section{Vertical distribution of H$_{3}^{+}$ density profiles}
\label{sec6a}
\begin{figure*}
 \includegraphics[width=12cm]{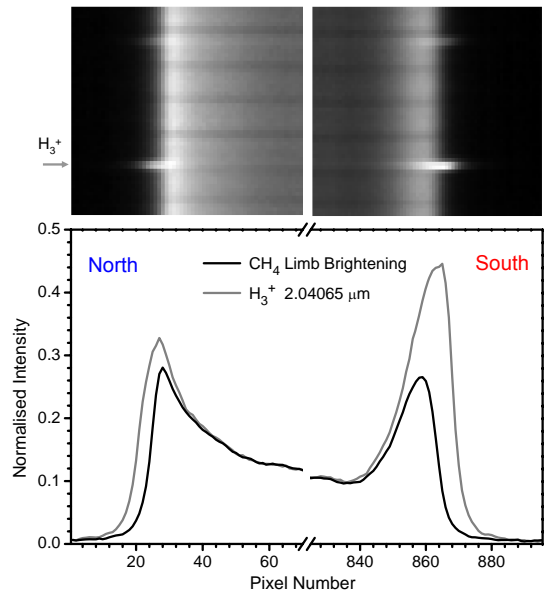}
  \caption{ {\it Top:} Two dimensional spectra in the region around the 2.0405~$\mu$m, H$_{3}^{+}$ line  {\it Bottom:} Intensity profiles corresponding to the H$_{3}^{+}$ line, and emission from the limb brightened continnum.}
\label{fig:8}
\end{figure*}

We use the second method applied in Section~\ref{sec5} to obtain spatial profiles of temperatures and column densities for the spectra derived from the
separate spatial pixels in the northern and southern auroral regions. Figure~\ref{fig:8} shows the two-dimensional spectra
in the regions under consideration and the corresponding intensity profiles, one across the brightest 2.0405~$\mu$m, H$_{3}^{+}$
line and the other across CH$_{4}$, limb brightened continnum. The peak pixel in the 2.0405~$\mu$m line profile is the same
as peaks in all other H$_{3}^{+}$ identifications in our high-resolution spectra.  

\begin{figure*}
 \includegraphics[width=13cm]{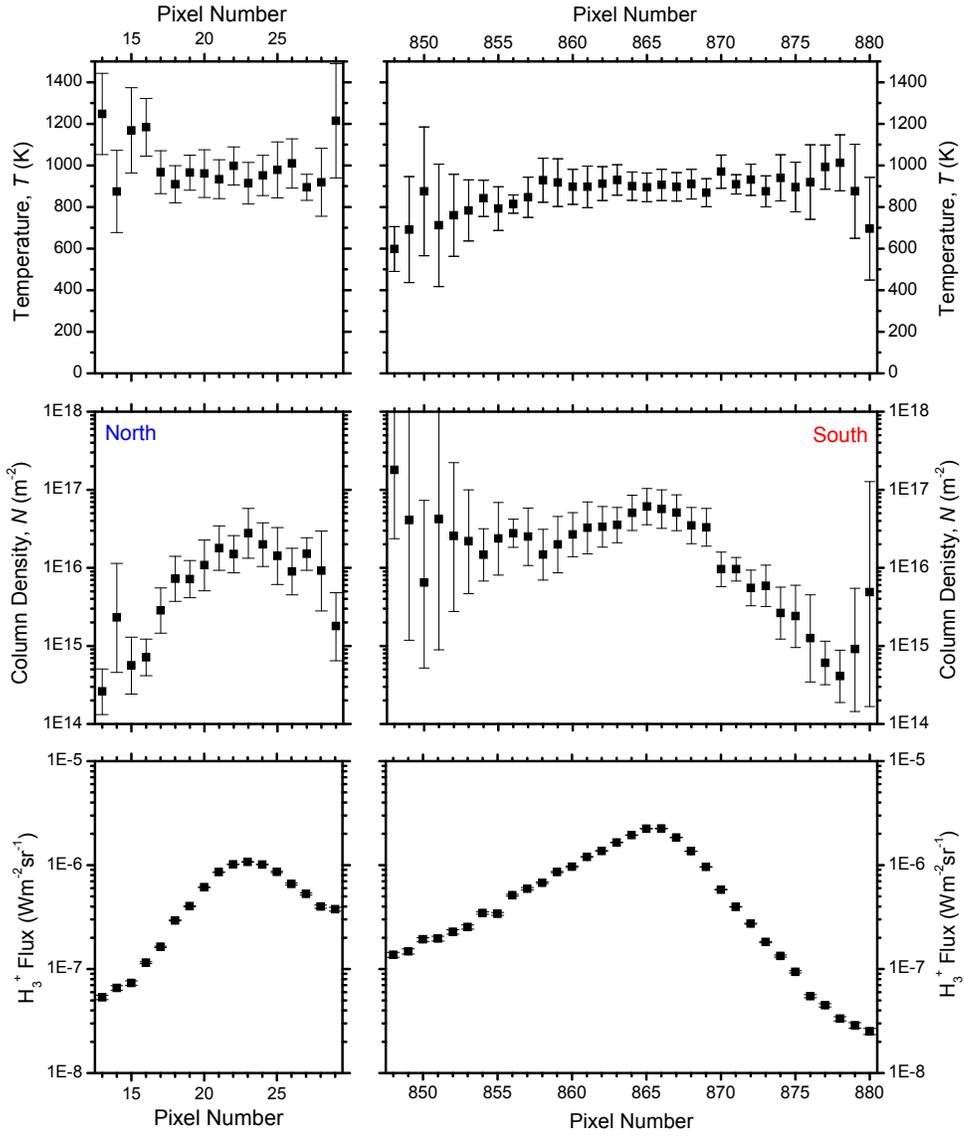}
  \caption{ 
Temperature ({\it top}) and  column density ({\it middle}) derived for spectra along single rows.  H$_{3}^{+}$ intensity profiles ({\it bottom}) are shown for the brightest emission line for the northern and southern aurora.
\color{black} 
}
\label{fig:9}
\end{figure*}

The fitted column densities, temperatures and spatial intensity profiles (Figure~\ref{fig:9}) show that the temperature remains
relatively unchanged through all selected pixels where auroral emissions dominate, while the column density variation seems to
correlate with the H$_{3}^{+}$ intensity profile. This was also noted in previously published observations by \citet{b38}.
However they also observed a strong anticorrelation in temperature and column density. \color{black} The higher column densities would be generated by the more energetic electrons penetrating deeper into the ionosphere where temperatures are lower. \color{black} In our data we see only weaker such effects at high altitudes in the northern region, and in the southern auroral spectrum away from the limb towards the center of the planet, where there is also a hint of temperature decreasing. 
\color{black} These differences, complicated by projection effects, could also be a result of increased noise in the regions
overlapping the planetary limb and in the upper-most atmosphere due to low signal-to-noise. \color{black} We intend to re-examine these trends in the new, currently scheduled observations. 
\color{black}

\begin{figure*}
 \includegraphics[width=12cm]{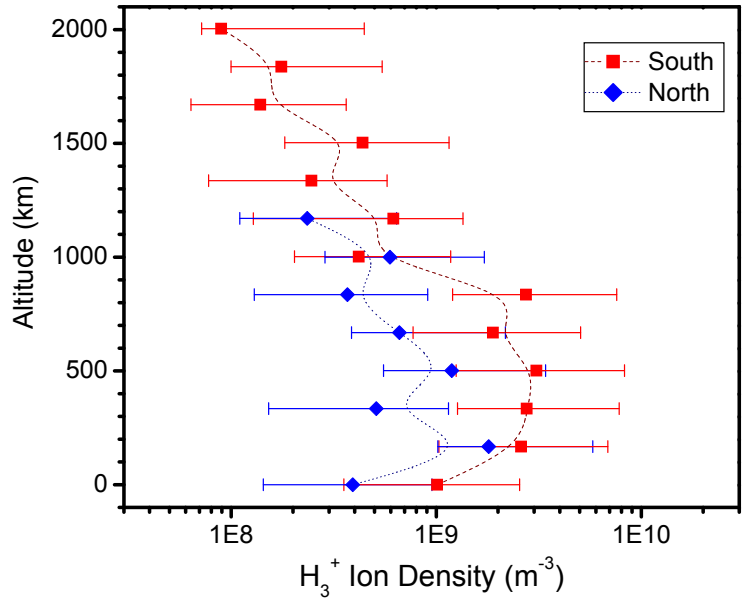}
  \caption{H$_{3}^{+}$ density profiles for the northern (blue) and southern (red) auroral regions. The spline fits to the profiles are shown with dotted lines. The errors were derived from the lower and upper bounds for column densities used in calculations.}
\label{fig:10}
\end{figure*}

Next we derive a vertical profile of ion density in both northern and southern aurora (Figure~\ref{fig:10}) by using the ``onion peeling" technique, where we divided the region above Jupiter's limb into concentric spheres of the size
corresponding to the physical dimension of our pixels (see e.g. \citet{b26,b38}). This method allows taking account of a
variation of column densities at different altitudes in the planetary atmosphere. 

\color{black}Figure~\ref{fig:10} shows higher ion densities in the southern region, associated with the main oval emissions than ion densities observed in the Io footprint region in the north. However both profiles appear rather similar with the broad peak at 300 - 800 km above the limb brightening altitude. 
 \color{black}
Similar analysis in \citet{b38} leads to significantly higher derived peak ion
densities and higher temperatures at the time of their observations in 2006 that reflects perhaps highly dynamic conditions in the ionosphere of the
planet (see the review of previous measurements in Section~\ref{sec7}).  
In common with \citet{b38} our data also show temperatures outside of the limb relatively constant while the column densities keep decreasing. However we do not observe the decrease of temperature in the dusk part of the auroral region (our northern spectrum). \color{black}This is most likely caused by the different orientation of our slit with respect to slit position in \citet{b38}, where we sample the emission along the auroral "curtain" in the vicinity of the Io spot, while they probe the region corresponding to the main oval in a direction perpendicular to the rotation axis. 
\color{black}

\section{H$_{2}$ quadrupole features in high and low-resolution spectra} \label{sec6} In the spectral region explored in
this study there are H$_{2}$ quadrupole $v$=1 to $v$=0 transitions that have been used in the past as probes of temperature
and density of the emitting material \citep{b27}. These transitions are thought to be excited due to either collisions or
fluorescent emission. Our low-resolution spectrum of the southern auroral region shows a prominent emission feature from the
1-0 S(1) rotational-vibrational transition at 2.1218 $\mu$m. We can also clearly identify the 1-0 S(2) line at 2.0338~$\mu$m
that is probably blended with the line at 2.0332~$\mu$m (that is visible in the high-resolution spectrum). At the low
wavelength edge of this spectrum the strong line of 1-0 S(3) at 1.9576~$\mu$m can be seen (Figure~\ref{fig:5}). However it is difficult to define
a continuum level around this line, which leads to large uncertainties in determining its intensity. \color{black} The low-resolution
spectrum of the northern region shows much weaker emission lines, which is not surprising as the slit did not extend beyond
the limb of the planet and we do not see bright emission associated with the main Io spot but possibly a weaker arc of the footprint projected on the limb. \color{black} Consequently both northern low and high-resolution spectra do not overlap spatially. 

We used the spectrum for the southern aurora, where the H$_{2}$ lines are clearly detected to derive estimates of
rotational temperature from the observed intensities of these three transitions. First we used Equation~\ref{eq:1} and
numerical values from Table~\ref{tab:13} to solve for T$_{rot}$ using 1-0 S(1) and S(2) lines that have well defined
continuum levels.  S(1) is an ortho and S(2) is a para state, therefore the assumptions about the ratio of both states has
to be considered. At high temperatures in Jupiter's ionosphere the H$_{2}$ para-ortho ratio can be reasonably assumed to be
0.25.  Using this ratio we calculated the rotation temperature, T$_{rot}$~=~575~$\pm$~245~K, where the errors were derived
from the estimate of 1 sigma deviation from the flux measurement of each line.

We repeated this procedure for the 1-0 S(1) and S(3) lines with the advantage of using both transitions in the same, ortho
state. However as we noted before the S(3) line is placed in the region of the residual absorption of the CH$_{4}$ band, and
it is impossible to derive precise level of continuum for this line. The 1--sigma error bar on the flux estimate of this line
is affected by its systematic component rather than a statistical deviation. The resulting temperature derived from the use
of these two transitions was T$_{rot}$~=~1025~$\pm$~350~K, which is consistent with  the temperature for 1-0 S(1) and S(2)
transitions within the range of the largest uncertainties derived in both cases.  If we use higher values of the para-ortho
ratio, this leads to lower temperature estimates from the 1-0 S(1) and S(2) pair.  \citet{b27} suggest that a ratio of 0.273
is a reasonable limit for the regions of the lower stratosphere where the in-situ measurements made 
with the Voyager's IRIS spectrometer \citep{b4} indicate 
\color{black}
temperatures of 170~K, and the auroral H$_{2}$ longitudinal position varies on time-scales that are much shorter than
possible for the ortho-para ratio to reach equilibrium. 

\begin{landscape}
\begin{table*}
\footnotesize
 \centering
 \begin{minipage}{18cm}
 \caption{Observed quadrupole  H$_{2}$ transitions in low-resolution spectra. The intensities are provided for the southern auroral region. J$^{\prime}$ is the rotational quantum number for the upper state, g$_{n}$ is the statistical weight of the nuclear spin, which depends on the spin alignment of both nuclei in H$_{2}$. At high temperatures g$_{n}$ of para states is 1 and ortho states is 3.}   
 \begin{tabular}{crccccccl}
    \hline \hline
  \textbf{Transition}&\textbf{Laboratory} &\textbf{Vacuum}& \textbf{Observed}  &\textbf{J$^{\prime}$}&\textbf{g$_{n}$}& \textbf{Upper }&\textbf{Einstein Coeff.}&\textbf{Measured}	\\
  
& \textbf{wavenumber}&			\textbf{wavelength}	&\textbf{wavelength}& &&\textbf{Energy Level}\footnote{The upper energy level is based on online data provided by D. L. DePoy where the rotational term value for upper state is derived by using constants from \citet{b28}. }&\textbf{A$_{ij}$ }  	&\textbf{Intensity}	 \\
\hline
&cm$^{-1}$	&   \AA&	\AA	& &&K&  s$^{-1}\times 10^{-7}$& W m$^{-2}$ sr$^{-1}$ \\
\hline
1-0 S(1) & 4712.905   & 21218 &21217 &3& 3&6956 &  3.47 &3.5$\pm$0.2$\times 10^{-6}$\\
1-0 S(2) & 4917.007  &     20338 &  20335    & 4& 1& 7584 &3.98&6.1$\pm$1.0$\times 10^{-7}$\\
1-0 S(3) & 5094.218 &  19576 & 19574 &5&3&8365 & 4.21& 1.9$\pm$0.6$\times 10^{-6}$ \\ 
 \hline
 \end{tabular} 
 \label{tab:13}
\end{minipage}
\end{table*}
\end{landscape}

\section{Previous measurements of temperatures and column densities of H$_{3}^{+}$ in auroral regions} \label{sec7}
\subsection{Temperatures} 
After the discovery of the near infrared 2$\nu_{2}$ overtone band of H$_{3}^{+}$ from 
Jupiter's aurora \citep{b5,b6}, the focus of subsequent observations quickly moved to the fundamental emissions that occur
at 3.5~$\mu$m. These are of an order of magnitude brighter and placed against the background of the dark limb of the planet.
The estimates of H$_{3}^{+}$ effective  temperatures were derived using both fundamental and overtone bands showing
rotational temperature values in the vicinity of 1000~K. 

The rotational temperature derived by \citet{b6} from the first identified lines in the overtone band was close to 1100~K under
the assumption of LTE. In 1990 the first measurement of the H$_{3}^{+}$ temperature from the fundamental band with the UKIRT
telescope gave a significantly lower value of 670~K \citep{b12}. Interestingly the overtone band at 2~$\mu$m was not
detected during that time, which implied, perhaps not surprisingly given observations of our Earth auroral emissions, that
auroral emissions can vary in time. \citet{b30} were able to detect both bands on the same night and found their relative
intensities indicated that H$_{3}^{+}$ vibrational levels 1 and 2 are populated as expected at the vibrational and
rotational temperature of 1100~$\pm$~100~K in thermal equilibrium. This also suggested that the dominant process in
H$_{3}^{+}$ is collisional excitation by H$_{2}$ and He rather than the resonant vibrational exchange with H$_{2}$. 
\citet{b31} tested the assertion about the H$_{3}^{+}$ emissions arising in thermal equilibrium. They calculated the
vibrational distribution of the molecule over the range of altitudes in the ionosphere assuming electron influx below levels
needed to produce the UV aurora. They concluded that at the peak of  H$_{3}^{+}$ emissions the radiation losses are almost
the same as collisional losses, which would lead to slight overpopulation of lower vibrational levels compared with the
Boltzmann distribution.  

Observations of the vertical and spatial distribution in auroral H$_{3}^{+}$ emission suggest that it peaks at around
0.2~$\mu$bar, which is equivalent to an altitude of 650 km above the cloud deck. \citet{b11}  showed
significant differences in rotational temperatures between northern (830~$\pm$~50~K) and southern (1000~$\pm$~40~K) auroral
regions at 4~$\mu$m. Subsequent imaging with narrow band filters showed that the auroral emission was clearly localised in ``hot spots''
\citep{b32}. \citet{b34} used a 2.7$^{\prime\prime}$ beam placed on the limb of Jupiter and confirmed significant 
differences between northern and southern temperatures of  H$_{3}^{+}$ emission. In addition, the temperature of  the
northern aurora appeared to drop by 200~K within one day. 

Later, more detailed mapping with higher spatial resolution revealed arc-like structures of strong emission surrounded by a
diffuse, low intensity area \citep{b33}. In fact \citet{b13} find that H$_{3}^{+}$ emission persists at all latitudes within
a few percent of auroral intensity.   A spectroscopic study by \citet{b25} at 3.47 and 4.00~$\mu$m allowed mapping of the
H$_{3}^{+}$ temperature and column density distribution over the entire disk of Jupiter with a resolution of about 10$^{o}$
in latitude. Maxima of temperatures are observed in the polar regions and are close to 1000~K. The highest estimate of
H$_{3}^{+}$ auroral temperatures came from the work of \citet{b14} who measured the ``hot band" transition R(3,4$^{+}$) at
3.94895~$\mu$m together with the fundamental band Q(1,0$^{-}$) line throughout the regions of auroral emission, allowing
derivation of  vibrational temperature, which ranged from 900 to 1150~K. Similarly high temperatures were derived by
\citet{b10} from the estimates of Doppler widths in observed lines from the collisional broadening. 

We noted in Section~\ref{sec5} the imaging of auroral zones using overtone band H$_{3}^{+}$ emission lines by \citet{b15} led
to estimates of both rotational and vibrational temperatures due to the presence of lines from the overtone and hot bands in
the spectral region examined. The average T$_{rot}$ of the $v_{2}=2$ level was determined to be 1170~$\pm$~75~K while 
T$_{vib}=960\pm50$~K for the $v_{2}=3$ level. Again such a difference suggests that the higher level $v_{2}=3$ is
underpopulated  as compared to expectation from the LTE. The rotational temperatures derived from our spectra in
Section~\ref{sec5} are slightly below 1000~K for both northern and southern auroras and are broadly consistent with the
range of values derived from the past observations of the overtone band.

\subsection{Column density} 
Even before the discovery of H$_{3}^{+}$ emission in Jupiter's ionosphere, its column density
was predicted from theoretical models that assumed production of H$_{3}^{+}$ by ionisation of H$_{2}$ and its destruction
due to recombination with electrons. \citet{b35} estimated the maximum concentration of H$_{3}^{+}$ above 1000~km above the
cloud deck and its column density between 6 and 10$\times10^{16}$~m$^{-2}$. 

The first measurements of column density from observations of the  $2\nu_{2}\rightarrow0$ band at 2~$\mu$m gave a value three
orders of magnitude lower than theoretical estimates, 2.8$\times10^{13}$~m$^{-2}$ \citep{b6}\footnote{The original value of
column density for $2\nu_{2}\rightarrow0$ band was corrected in \citet{b11}.}. \citet{b11} estimated column density from
$\nu_{2}\rightarrow0$ band observations under the assumption of thermal equilibrium and also found it lower than theoretical
predictions. They measured  3.3$\times10^{14}$~m$^{-2}$ in the southern auroral region. In two observations three days apart
they obtained 1.2 and 2.4$\times10^{14}$~m$^{-2}$ in the northern auroral region. 

In a model of vibrational distribution of H$_{3}^{+}$, \citet{b31} find its peak to be slightly higher than 650 km above the
cloud deck and derive maximum column densities at the peak to be consistent with the observations. In addition, their model
shows column densities of the overtone band systematically lower than the fundamental band as a function of height. This
also appears to be supported by 
previously published 
\color{black}observations.  Since their model suggested a deviation from local thermal equilibrium at the
peak of H$_{3}^{+}$, any measurements that assumed LTE should be treated as lower limits to actual levels of its column
density.

Notably the first measurements of column densities were averages over several thousands of kilometres of Jupiter's area,
because the spectra were obtained with low spatial resolution.   Observations of the full planetary disk at 3.47 and
4.0~$\mu$m with close to 10$^{4}$ km resolution show a significant column density of H$_{3}^{+}$ at above
0.1$\times10^{16}$~m$^{-2}$ that is also measured at the equatorial latitudes \citep{b25}. The subsequent  mapping
observations \citep{b14, b15} revealed the rather complex and time variable distribution of H$_{3}^{+}$ within auroral
regions. \citet{b14} who scanned the northern auroral region in two emission lines of H$_{3}^{+}$ over five nights, find
significant variation in average column density in the range from 0.58 to 1.1 $\times10^{16}$~m$^{-2}$. The most detailed
maps of column density in both northern and southern auroral regions were derived in ($2\nu_{2}\rightarrow0$) 2 $\mu$m band
with the extreme values close to $10^{19}$~m$^{-2}$ in the arc-like feature of the northern aurora. 

Our measurements of N(H$_{3}^{+}$) with 3.1$\times10^{16}$~m$^{-2}$ and 5.1$\times10^{16}$~m$^{-2}$ for the northern and
southern auroras respectively, are consistent with previous observations that had comparable spatial resolution to what can
be achieved with the Gemini telescope.

\section{Discussion} \label{sec8} Our detections of auroral emission lines and subsequent measurements of rotational
temperature, T$_{rot}$ and column density of the H$_{3}^{+}$ emissions are just one spatial and temporal snapshot that shows
only one aspect of the energy exchange processes in Jupiter's ionosphere. 

Dedicated observations of auroral emission in infrared bands with 8-m class telescopes like Gemini and larger are still
quite rare despite the clear advantage of superb spatial resolution of the Jovian disk achievable from the ground. Two
recently published studies show such results of observations taken with the largest telescopes. The Keck II telescope
mapping of emissions in L-band \citep{b38} with NIRSPEC and observations with the IRCS spectrometer on the Subaru telescope in
K-band \citep{b26} allowed derivation of vertical ion density and emissivity profiles based on echelle spectra of
H$_{3}^{+}$ taken with resolution better than the R~=~18000 observations reported here.  

\subsection{Spectroscopy of auroral emissions} 
Long slit observations provide typically better throughput and less chance of
spurious lines from overlapping orders as compared with multi-dispersing designs. This is particularly useful for examining
the low intensity features seen in our spectra, obtained with the highest resolution of GNIRS. The infrared windows in K and
L bands, where the methane absorbs almost completely any thermal emission from the planet, are naturally chosen for
observation of auroral emission. 

However telluric absorption due to molecules in the Earth's atmosphere requires very careful calibration that is often too
difficult using the traditional method of subtracting a spectrum of the standard star from the target object. In addition,
the standard star's spectral type has to be closely matched to solar G2V to avoid stellar lines imprinting on the target
spectrum.  This method can leave spurious residual features in spectra, as seen for example in Figure 4 of  \citet{b26},
where telluric absorption of CO$_{2}$ bands around 2.0 and 2.06~$\mu$m in Figure~\ref{fig:3}(b) were overcorrected.  The
intensity of spurious emission lines in their spectra is sometimes higher than the weak lines assigned to H$_{3}^{+}$, which
can lead to misidentifications. The ATMOF method \citep{b16} used for our spectral calibration deals more precisely with the
effects of telluric absorption. This allows us to study any weak emissions, such as lines visible in ``zoomed in''
Figure~\ref{fig:6}. 
\color{black}
\subsection{The H$_{3}^{+}$ spatial distribution and correlation with H$_{2}$ emission } 
\begin{figure*}
\includegraphics[width=11cm]{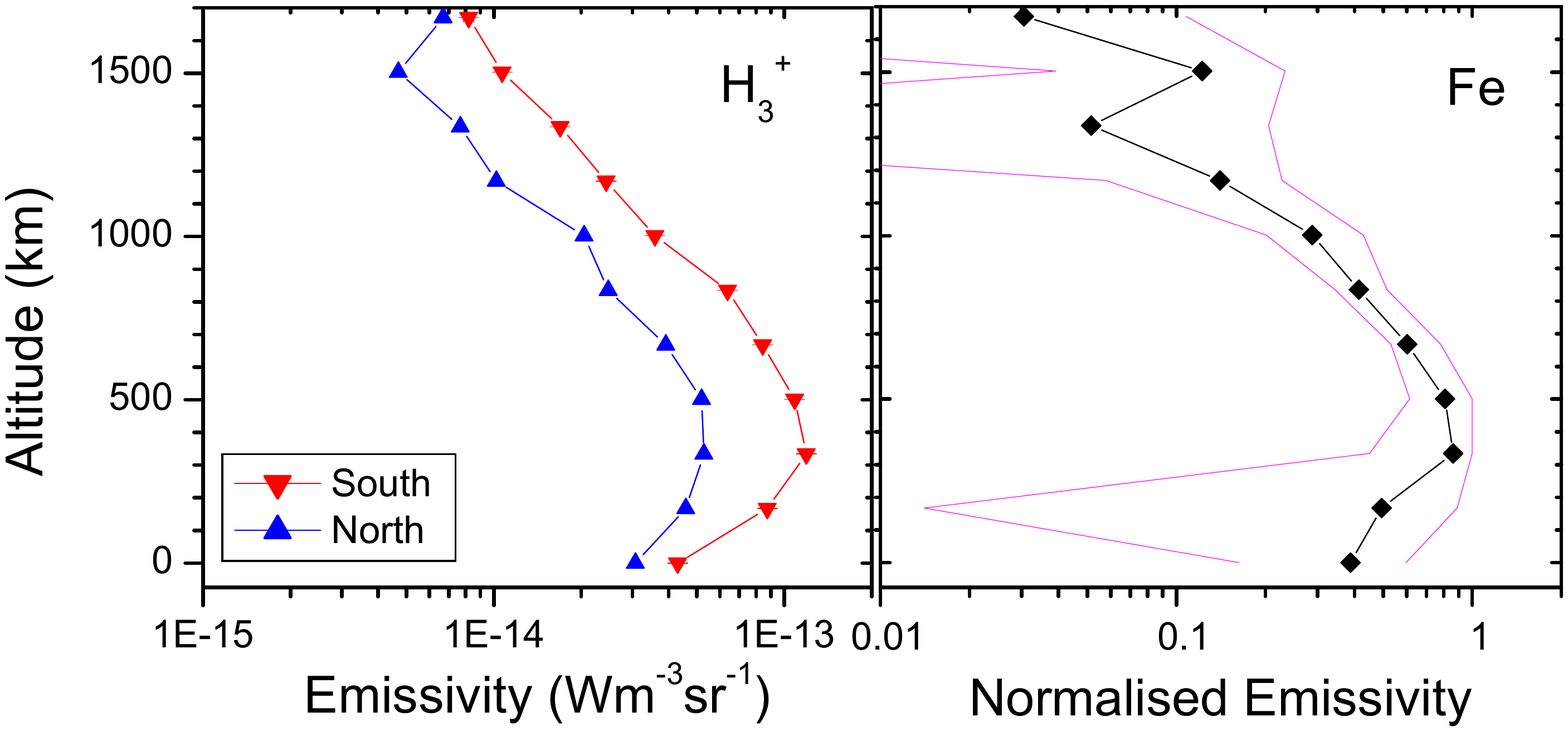}
 \caption{ 
{\it Left:} Emissivity profiles of the  H$_{3}^{+}$ 2.0405~$\mu$m line for the southern and northern auroral region. {\it Right:} The average normalised emissivity profile of two strongest lines Fe~I 2.0234 and~2.0284~$\mu$m combined for both regions. The upper and lower bounds shown in the figure represent the highest and lowest normalised emissivities amongst the four lines. 
 \color{black}
}
\label{fig:13}
\end{figure*}

\begin{figure*}
\includegraphics[width=12cm]{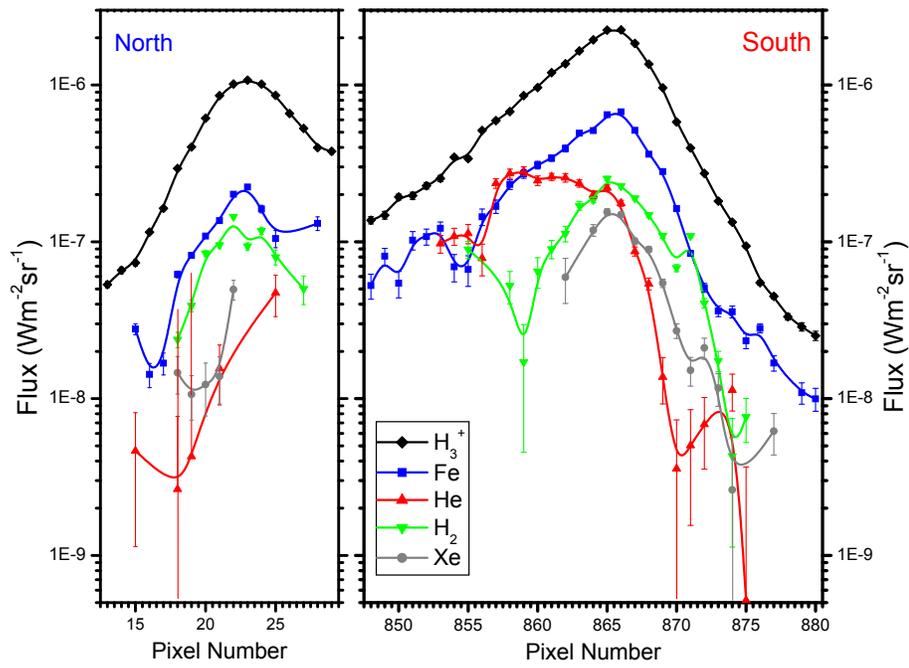}
 \caption{Intensity profiles of the emission lines that belong to assigned atomic and molecular species. H$_{3}^{+}$ -- 2.0405~$\mu$m, Fe~I -- 2.0231~$\mu$m, Xe~I -- 2.0268~$\mu$m, H$_{2}$ -- 2.0338~$\mu$m and He~I -- 2.0587~$\mu$m are shown in the plots. The lines connecting data are the spline fits to the profiles.   }
\label{fig:11}
\end{figure*}

\color{black}
Our resulting rotational temperatures and column densities of
H$_{3}^{+}$ for the small portions of Jupiter aurora covered by the GNIRS spectra are similar to the values derived for well
spatially resolved maps of auroral emissions  \citep{b25,b15}. A single position of the slit provides very limited
information on the vertical distribution of the peak of H$_{3}^{+}$ emission. 
The vertical emissivity profile for H$_{3}^{+}$ was derived by application of the ``onion peeling" technique for the northern and southern regions (Figure~\ref{fig:13}). The peak of emissivity is visible within the broad layer centered at 400 km above the peak of the limb brightening discussed previously, which is defined as the zero altitude. We attempted to obtain vertical emissivity profile for H$_{2}$ from the line at 2.0334~$\mu$m as well. It suggests a more uniform emissivity distribution over the region of altitudes covered in Figure~\ref{fig:13}. The difference is reasonably well visible in the intensity profiles of H$_{2}$ and H$_{3}^{+}$ emission along the slit for both auroral regions (Figure~\ref{fig:11}), which is not surprising considering the chemistry of both molecules. 
\color{black}

The formation mechanism of H$_{3}^{+}$ involves collisions of H$_{2}$ with ionised molecular hydrogen. The produced ion is
vibrationally excited and this energy is re-radiated in infrared spectral lines such as presented in this report. This radiative
cooling is an important process in energy balance of the ionosphere, together with radiative cooling by CH$_{4}$ and
C$_{2}$H$_{2}$ at the lower atmospheric heights.  In the upper ionosphere the H$_{3}^{+}$ ion is destroyed by interaction with an electron in
dissociative recombination, a mechanism that adds heat into the ionosphere. The lifetime of the ion depends on the
density of electrons which seem to peak at low altitudes (below 1000~km) during dusk, and at 1500 to 2000~km at dawn \citep{b70}.  The corresponding lifetimes of H$_{3}^{+}$ ion are of the order of 10 to 100 sec. In
the Jovian aurora the main source of electrons available to ionise hydrogen is the solar wind \citep{b32} and the ions formed from gases emitted by volcanic activity on Io that are trapped in the planetary magnetic field
and spiral down towards its surface along field lines. 
%
Our emissivity profiles for H$_{3}^{+}$ ion in the northern and southern regions that correspond to dusk and dawn limbs respectively, show peaks at significantly different altitudes, which may be related to different electron densities in these regions. Another reason for this difference could be the geometry of the emissions in both regions. \color{black}In the south the slit covers more of the auroral oval and our line of sight cuts through longer extents of brighter emission, while in the north we do not sample emission associated with the main auroral oval but a slightly weaker emission from the Io footprint. 
The statistical altitude of the peak UV emission for the main Io spot was suggested to be 900$\pm$125 km from the 1 bar level \citep{b77}, which is significantly higher than our peak of the volume emissivity at $\sim$650 km for H$_{3}^{+}$ (assuming the limb brightening position at 200 km above 1 bar level). This difference could be explained by our observations corresponding to higher energies of precipitating electrons than suggested in \citet{b77}.
\color{black} 

The quadrupole transitions of H$_{2}$ in K-band were associated with Jupiter's aurora before discovery of the overtone
H$_{3}^{+}$ in that region \citep{b5}. There is still some debate over the similarity in spatial distribution of both
molecules. Although spatially resolved ultraviolet observations with the HST \citep{b42} were first to note very close
correspondence of H$_{2}$ to H$_{3}^{+}$ emissions, the maps in \citet{b15} seem to show significant differences in the
relative intensity of emission for both molecules. Proposed reasons for this could be a variable ortho-to-para ratio in
H$_{2}$ distribution or dynamic changes due to supersonic winds \citep{b43,b41} Considering vertical distribution of H$_{2}$
and H$_{3}^{+}$, the model by \citet{b31} suggests the peak intensity of their emission at the same height, close to 600~km
above the cloud deck. This conclusion was supported by observations of emissivity profiles undertaken by \citet{b26}. 
Our emissivity profile derived from only one weak line of H$_{2}$ is not sufficiently reliable to reach firm conclusions about the altitude differences between both species. However the distribution of intensities along the slit in Figure~\ref{fig:11} suggests the H$_{2}$ emission peaks slightly below the H$_{3}^{+}$, and the emissivity is reduced less rapidly with increasing altitude compared with the H$_{3}^{+}$ ions.  \color{black}
\subsection{The LTE considerations}  
Temperatures derived from infrared emission lines of  H$_{3}^{+}$ are associated with
the transitions present in the observed portions of its ro-vibrational spectrum. Our data allows only determination of the
rotational temperature, because only  transitions from the single $2\nu_{2}\rightarrow0$ overtone band are identified in the
spectral region between 2.01 and 2.07~$\mu$m. A comparison of line intensities for two different vibrational bands is
required to assess whether T$_{rot}$ is also a thermal temperature of H$_{3}^{+}$ in the ionosphere, which means that the
relative populations at different vibrational levels follow a Boltzmann distribution (are in thermal equilibrium). 

In a previous study \citet{b30} measured the vibrational temperature from line intensities of both fundamental and overtone
bands and found it consistent with the rotational temperatures, which suggests that H$_{3}^{+}$ is in the LTE condition. In
addition the translational temperatures measured from resolved line shapes and compared with rotational temperatures 
\citep{b10} suggest that T$_{rot}$ represents the kinetic temperature in Jupiter's ionosphere. However this assertion is not
entirely correct since models \citep{b31,b39} show that at the altitude of the strongest H$_{3}^{+}$ emissions the rate of
radiative cooling and collisional de-excitations becomes similar for the $2\nu_{2}$ and $\nu_{2}$ vibrational levels.
Although the ratio of populations at these levels satisfies the Boltzmann distribution, the ground level is excessively
populated by radiative cooling. This effect becomes even more important at higher altitudes and it affects cooling
efficiency of H$_{3}^{+}$ in the upper ionosphere.

\citet{b44} developed a two-stream electron transport model which links the spectrum of incident electrons with distribution
of vertical temperature in the ionosphere. Since  H$_{3}^{+}$ thermal emissions are affected mainly by the soft-energy
electrons (3~keV), they conclude that they cannot be used to derive the entire energy driving the auroral display. Thermal
emissions from  H$_{2}$ and C$_{2}$H$_{2}$ should be used to help constrain the high-energy end of the electron flux
component. 
\citet{b26} modeled the altitude profiles for one of the H$_{3}^{+}$ lines from $2\nu_{2}$ vibrational levels under assumption of the LTE and used a broader distribution of electron fluxes than assumed in \citet{b44}.  The resulting emissions peaked at altitudes above 1000~km -- much higher than the 400--800~km suggested by their observations. Similarily our data suggest lower peak altitudes around 600~km. \citet{b26} showed that relaxing the LTE assumption could bring the peak altitudes in closer agreement with their estimates from observations (Figure~\ref{fig:10}).  
  
Mapping observations \citep{b15,b14} suggest a clear spatial correlation of emission line intensity with column densities of
H$_{3}^{+}$ and only marginal correlation with temperatures. 
In Figure~\ref{fig:9} temperature profiles fitted to our data across the slit are relatively constant while column densities   are correlated with H$_{3}^{+}$ emission distribution. Since H$_{3}^{+}$ acts like a very efficient thermostat the increase in temperature leads to a faster cooling rate, which in turn moderates overall changes in temperature. Therefore observed line intensities are more representative of column densities in the upper ionosphere and are most sensitive to the H$_{3}^{+}$ distribution. It is important to note that depending on the amount of deviation from LTE conditions the fitted temperatures underestimate the actual temperature and the ion density derived from the column density, N(H$_{3}^{+}$)  \citep{b39}.  
\color{black}
\subsection{Additional emission lines in high-resolution spectra.}   
\label{sub8.4} 
Figure~\ref{fig:4} shows that all 
H$_{3}^{+}$ lines, which have transitions in the $2\nu_{2}\rightarrow0$ overtone band in this spectral region, were easily
identified. However these spectra show a few more emission lines, most of them weak but also a few relatively strong that do
not correspond to any confirmed transitions of H$_{3}^{+}$ \citep{b21}. Beside the expected H$_{2}$ quadrupole S(2) line at
2.0334~$\mu$m and the resonant He I at 2.0587~$\mu$m , there are strong features at 2.0231, 2.0284, 2.0289, 2.0332, 2.0635 and 2.0683~$\mu$m that were missing
identification. We examined the high-resolution spectra for potential sources of spurious lines and hypothesised that some
of these emissions could be due to atomic species and searched the NIST database \citep{b22} for appropriate
identifications. 

Voyager 1 {\it in situ} observations of Jupiter's magnetosphere detected a torus-shaped cloud of heavy ions and energetic
neutral atoms (mainly oxygen, sodium and sulfur) that originate from the volcanic activity of Io \citep{b47,b45}. At first a
precipitation of these ions into high latitudes along magnetic field lines was thought to provide enough energy to excite
Jupiter's aurora \citep{b46}; this idea was subsequently disputed by \citet{b48}. Ionised atoms are typically observed by
their X-ray emissions \citep{b49}. Recent observations of Jovian transmission spectrum by \citet{b40} reveal the presence of
neutral sodium in the upper atmosphere of the planet. They suggest its origin from either cometary impacts, as observed
during the collision of comet Shoemaker-Levy 9 (SL9) in 1994 \citep{b50}, or due to deposition from Io into a toroidal area
around the planet.

There are 377 different atomic transitions listed in the NIST database within the range of our high-resolution spectrum. We
examined lines from different atomic species by use of their relative intensities (keeping in mind that these are rather
uncertain parameters). If the strongest transitions of a given (neutral or ionised) atom were clearly not visible in our
spectrum, this would make this species an unlikely identification, even if some lines matched its weaker transitions. We
found the most consistent matches with Fe~I transitions and also identified a strongest emission of Xe I in this region.
Finding matches with Fe perhaps should not be surprising, since its abundance is high in the interstellar medium. Iron also
has a complex and rich energetic structure. In our spectral region Fe~I has close to a third of all transitions listed.
Although this could be a reason why we are more likely to match our lines with the Fe transitions, the transitions of
ionised iron are almost twice as common as Fe~I, so we would expect them to match lines in our spectra more often than
transition for neutral iron if the matches were random. In such a case we would also expect matches with many other atomic
species. There is no strong correlation between the intensities listed in the database and those observed in our spectrum,
which is a result of either incorrect assignments or the very different conditions under which the lines listed in the
database were observed compared with the conditions of Jupiter's atmosphere.     

Figure~\ref{fig:6} shows a zoomed-in spectrum in order to focus on weaker lines. The upper panel shows the stick spectrum of
Fe~I transitions, which are aligned with our high-resolution spectrum. We marked lines in black when there is a good match
with transitions listed in Table~\ref{tab:2} in bold. The dotted lines are centred on the peaks of the lines that are the
closest to known Fe~I transitions, the weakest of them are clearly just above our signal-to-noise level. In
Table~\ref{tab:2} we comment on the possible problems with the individual assignments.   

In Figure~\ref{fig:11} we plotted the intensity profiles of lines of different species including our Fe~I assignments. All
of them are similar in shape to H$_{3}^{+}$ profile with a notable exception of the resonant He~I line with a broader intensity
peak that appears much closer to the limb of the planet. Of course a possible reason for the similarity of H$_{3}^{+}$ and Fe~I
profiles could be incorrect assignment of unknown H$_{3}^{+}$ transitions. However such unknown transitions are likely to be
considerably weaker than observed unassigned lines in our spectrum. 

In the right panel of Figure~\ref{fig:13} we show the averaged emissivity profile for the Fe~I identifications obtained with the onion peeling method. Initially we plotted the strongest two Fe lines from each hemisphere, and discerned no significant differences in their emissivity-altitude profiles. Subsequently the data for these lines (2.0234 and 2.0283~$\mu$m) for northern and southern regions have been merged as follows. For each line the data is normalised to the altitude of greatest emissivity. The average of the four normalised lines at each altitude has then been plotted in Figure~\ref{fig:13}. The upper and lower bounds shown in the figure represent the highest and lowest normalised emissivities amongst the four lines. The normalised Fe emissivity has been plotted over the same number of decades as the corresponding H$_{3}^{+}$ emissivity in the left panel of the figure for easy comparison of their rates of change. Profiles for both species appear to have the strongest emissivity at similar altitude.  
\color{black}

The only published observations of iron emissions in Jupiter's atmosphere were the HST/FOS UV spectrum taken after the
impact of the SL9 comet (Figure 4 in \citet{b50}) and the spectrum of optical emission from two multiplets at 0.445 and
0.52~$\mu$m centred on the comet impact sites \citep{b37}. Solar fluorescence was suggested as a mechanism of the UV
emissions rather than a collisional excitation that is less efficient in higher levels of Jupiter's atmosphere, however
optical emissions were assumed to arise due to thermal excitation of the cometary material.

Neutral iron emission due to fluorescence was observed in the Earth airglow  during twilight at 0.3860~$\mu$m by \citet{b51}. A
possible source of Fe~I at the heights close to 100~km in the Earth's atmosphere is suggested to be meteor ablation.   
Similar processes could lead to periodic occurrence of neutral and ionised iron in the high levels of Jupiter's atmosphere.
The infrared emission of neutral Fe~I is produced by transitions at high excitation states. Such highly excited atoms are
thought to originate from a recombination of Fe ion \citep{b52}, which could be important in the auroral regions of Jupiter
with the enhanced electron density. 

\citet{b72} and later \citet{b73} derived a model of the ion densities for the products of meteoric albation in Jupiter's atmosphere. They find a layer of ions that peaks at the altitudes of around 400~km above the level of 1 bar pressure. This peak is dominated by the contribution of Mg~II and Fe~II with the ion density close to $10^{10}$~m$^{-3}$. The total density of ions remains similar at the higher levels of the ionosphere but it is dominated by the H$_{3}^{+}$ contribution, while the Fe~II ion density drops steadily to negligible value at altitudes close to 1000~km. \citet{b72} estimated production rate of the neutral Fe at the higher altitudes to be orders of magnitude larger than Fe~II. The emissivity of neutral Fe~I from the lines detected in our spectra in Figure~\ref{fig:13} peaks around 600~km above 1 bar pressure, which is higher than suggested by the model for ionised content of the ionosphere. In future investigations we intend to model the characteristic de-excitation times for these possible Fe~I transitions, which will help us to derive their density profiles and compare them with estimates for the ionised Fe~II distribution. 
\color{black}

\section{Conclusions} 
\label{sec9} 
\color{black}In summary, we obtained high- and low-resolution spectra of Jupiter's two auroral regions with GNIRS
on the Gemini North telescope. The northern region is likely to be associated with the main spot in the Io footprint and the southern region samples the area of the main oval. We identified strong emission lines from the H$_{3}^{+}$ overtone band and H$_{2}$ quadrupole emissions in both low- and high-resolution spectra. \color{black} We also tentatively suggested assignment
of  weaker lines with atomic transitions of Fe~I, He I and Xe I. Our use of ATMOF fitting allowed a superior removal of telluric
absorption from CO$_{2}$ bands. This improves our confidence in fitting and assigning weaker lines in the spectra. 

Transitions of H$_{3}^{+}$ were used to derive rotational
temperature and column density for both northern and southern auroral regions, and we found that our values are broadly
consistent with previously published measurements. 
\color{black}
We derived vertical emissivity and H$_{3}^{+}$ ion density profiles of both auroral regions which includes an estimate of the altitude for the emission peak associated with the main oval of the Io footprint.
\color{black}
Since we were not able to determine if the rotational temperature derived here is a true kinetic temperature due to
observations of only one vibrational band of H$_{3}^{+}$ and we also had very limited spatial information about observed
emissions, we intend to make 
additional long slit auroral mapping observations with the GNIRS at L and K band almost simultaneously to
constrain the parameters derived from H$_{3}^{+}$. These observations are planned during the Juno mission flight over polar regions in order to measure electron density distribution together with associated UV and infrared emissions. 

New spectra obtained in the spatial mapping mode will also help us to determine the identifications and provide  emissivity profiles of lines that have not yet been assigned. 
Our spectra will extend over a wider range of wavelengths that will allow us to search for the stronger lines of neutral or ionised atomic ions and to confirm of disprove our tentative observations of neutral iron lines.  
This extended spectral coverage will also allows us to understand the spatial distribution of the weak He~I emission. \color{black}
 
Although Jovian aurora have been studied by different methods that provided clues about origin, physical
conditions, heating/cooling processes, energy transport and LTE balance; many results are still debated. Monitoring of
auroral emissions is even more important at the time when the Juno mission is starting to release the first data for {\it in
situ} study of Jupiter's upper atmosphere, magnetosphere and auroras. Continuation of ground-based studies is intended
to place Juno's observations in a global context. Jupiter's auroral emission provides a laboratory from which we can draw
analogies for possible observations of such emissions in the exoplanets, brown dwarfs and late type stars like the recently
observed aurora detected in the M8.5 dwarf, LSR J1835+3259 \citep{b53}.

\section*{Acknowledgments}
This work is based on observations (GN-2011B-Q-1) obtained at the Gemini Observatory, which is operated by the 
Association of Universities for Research in Astronomy, Inc., under a cooperative agreement 
with the NSF on behalf of the Gemini partnership: the National Science Foundation 
(United States), the National Research Council (Canada), CONICYT (Chile), the Australian 
Research Council (Australia), Minist\'{e}rio da Ci\^{e}ncia, Tecnologia e Inova\c{c}\~{a}o 
(Brazil) and Ministerio de Ciencia, Tecnolog\'{i}a e Innovaci\'{o}n Productiva (Argentina).
We also wish to thank Professor Michael Burton for his helpful discussion and advice that benefitted this paper. We wish to thank the anonymous referees for the follow-up research suggestions that helped us to improve the paper.




\end{document}